\documentclass{IEEEtran} 
\IEEEoverridecommandlockouts

\usepackage{graphics}
\usepackage{epsfig}
\usepackage{times}
\usepackage{amsmath}
\usepackage{amssymb}
\usepackage{graphicx}
\usepackage{epstopdf}
\usepackage{epsfig}
\usepackage{enumerate}
\usepackage[noadjust]{cite}
\usepackage{color}

\usepackage{makecell}
\usepackage{arydshln}
\usepackage{stmaryrd}
\usepackage{stackengine}
\usepackage{threeparttable}
\usepackage{slashbox}
\usepackage{caption}
\usepackage{algorithm}
\usepackage{algorithmicx}
\DeclareCaptionFormat{myformat}{#3}
\newtheorem{Theorem}{Theorem}
\newtheorem{Lemma}{Lemma}
\newtheorem{Definition}{Definition}
\newtheorem{Remark}{Remark}
\newtheorem{Assumption}{Assumption}

\title{\LARGE \bf Algorithm-Level Confidentiality for Average Consensus on Time-Varying Directed Graphs}

\author{Huan~Gao,~\IEEEmembership{Member,~IEEE,} and Yongqiang~Wang,~\IEEEmembership{Senior~Member,~IEEE}
\thanks{\bf This paper has been accepted to IEEE Transactions on Network Science and Engineering as a regular paper. Please cite this paper as: Huan Gao and Yongqiang Wang, ``Algorithm-Level Confidentiality for Average Consensus on Time-Varying Directed Graphs,'' in IEEE Transactions on Network Science and Engineering, doi: 10.1109/TNSE.2022.3140274.}
\thanks{Part of the results was presented at 2018 IEEE Conference on Communications and Network Security (CNS) \cite{Huan2018CNS}. The work was supported in part by the National Science Foundation under Grants ECCS-1912702 and CCF-2106293.}
\thanks{Huan Gao was with the Department of Electrical and Computer Engineering, Clemson University, Clemson, SC 29634, USA. He is now with the School of Automation, Northwestern Polytechnical University, Xi'an 710129, China ({\tt\small email: huangao@nwpu.edu.cn}). The work was done when Huan Gao was with Clemson University.}
\thanks{Yongqiang Wang is with the Department of Electrical and Computer Engineering, Clemson University, Clemson, SC 29634, USA ({\tt\small email: yongqiw@clemson.edu}).}
}

\begin{document}	
\maketitle

\begin{abstract}
Average consensus plays a key role in distributed networks, with applications ranging from time synchronization, information fusion, load balancing, to decentralized control. Existing average consensus algorithms require individual agents to exchange explicit state values with their neighbors, which leads to the undesirable disclosure of sensitive information in the state. In this paper, we propose a novel average consensus algorithm for time-varying directed graphs that can protect the confidentiality of a participating agent against other participating agents. The algorithm injects randomness in interaction to obfuscate information on the algorithm-level and can ensure information-theoretic privacy without the assistance of any trusted third party or data aggregator. By leveraging the inherent robustness of consensus dynamics against random variations in interaction, our proposed algorithm can also guarantee the accuracy of average consensus. The algorithm is distinctly different from differential-privacy based average consensus approaches which enable confidentiality through compromising accuracy in obtained consensus value. Numerical simulations confirm the effectiveness and efficiency of our proposed approach.
\end{abstract}


%%%%%%%%%%%%%%%%%%%%%%%%%%%%%%%%%%%%%%%%%%%%%%%
\section{Introduction}

Averaging consensus is an important tool in distributed computing. For a network of $N$ agents interacting on a graph, average consensus can enable the states of all agents to converge to the average of their initial values through local interactions between neighboring agents.

Recently, average consensus is finding increased applications in load balancing \cite{boillat1990load, cybenko1989dynamic}, network synchronization \cite{lynch1996distributed}, distributed information fusion \cite{scherber2004locally, xiao2005scheme}, and decentralized control \cite{olfati2004consensus, ren2005consensus}. To ensure all agents converge to the average value of their initial values, conventional average consensus approaches require individual agents to exchange explicit state values with their neighbors. This results in the disclosure of sensitive state information, which is sometimes undesirable in terms of confidentiality. In fact, in many applications such as the smart grid, health-care or banking networks, confidentiality is crucial for promoting participation in collaboration since individual agents tend not to trade confidentiality for performance \cite{mangasarian2011privacy, hoenkamp2011neglected, lou2017privacy}. For instance, a group of people using average consensus to reach a common opinion may want to keep their individual opinions secret \cite{tsitsiklis1984problems}. Another typical example is power systems in which multiple generators have to reach agreement on cost while maintaining their individual generation information confidential \cite{zhang2011incremental}.

To achieve confidentiality in average consensus, recently results have started to emerge. A commonly used confidentiality mechanism is differential privacy from the database literature \cite{huang2012differentially, huang2015differentially, nozari2015differentially, nozari2017differentially, katewa2017privacy, gao2019differentially, ye2019differentially} (and its variants \cite{kefayati2007secure, wang2018privacy}) which injects independent (and hence uncorrelated) noises directly to agents' states in order to achieve confidentiality in average consensus. However, the use of independent noises on the states in these approaches prevents converging to the exact average value \cite{wang2017differential}. To improve consensus accuracy, which is crucial in cyber-physical systems and sensor networks, \cite{ he2018distributed, he2019consensus, he2018privacy, mo2017privacy, charalambous2019privacy, gupta2019statistical, pilet2019robust, manitara2013privacy} inject carefully calculated {\it correlated} additive noises to agents' states, instead of independent (and hence uncorrelated) noises used in differential-privacy based approaches. (A similar approach was proposed in \cite{duan2015privacy} to achieve maximum consensus.) However, these prior works only consider average consensus under balanced and static network topologies. Different from injecting noises to agents' states in the aforementioned approaches, \cite{altafini2019dynamical} employed carefully designed mask maps to protect the actual states. Observability based approaches have also been reported to protect the confidentiality of multi-agent consensus \cite{pequito2014design, ridgley2019simple, alaeddini2017adaptive}. Its idea is to design the topology of interactions such that the observability from a compromised agent is minimized, which amounts to minimizing the ability of the compromised agent to infer the initial states of other agents. Recently, encryption based approaches have been proposed to protect the confidentiality by encrypting exchanged messages with the assistance of additive homomorphic encryption \cite{ kishida2018encrypted, hadjicostis2020privacy, fang2018secure, ruan2019secure}, with the price of increasing computation and communication overhead. Another confidentiality approach was proposed in \cite{wang2019privacy} where each agent's confidentiality is protected by decomposing its state into two sub-states. However, \cite{wang2019privacy} relies on undirected interactions and is inapplicable to time-varying directed graphs considered in this paper.

This paper addresses the confidentiality of average consensus under time-varying directed graphs that are not necessarily balanced. Since push-sum based average consensus approaches do not require balanced topologies, we build our confidential average consensus algorithm on the push-sum approach. More specifically, to protect the confidentiality of the initial states of participating agents, in the first several iterations, we let agents send random values instead of their actual states to obfuscate their initial values. Of course, to guarantee the accuracy of average consensus, we have to judiciously design the state-update rule such that the randomness added in the first several iterations does not affect the final convergence result. Different from approaches injecting correlated additive noises directly to agents' states, our approach adds independent (and hence uncorrelated in time) randomness directly to the average consensus dynamics, which makes it applicable to time-varying directed graphs. Compared with our prior work in \cite{ruan2019secure} which employs homomorphic encryption to preserve confidentiality and \cite{wang2019privacy} which protects each agent's confidentiality by decomposing its state into two sub-states, this paper proposes a different approach that enables confidentiality by judiciously adding randomness in interaction dynamics. More importantly, both \cite{ruan2019secure} and \cite{wang2019privacy} rely on undirected interactions and hence are inapplicable to time-varying directed graphs considered in this paper. Some of the results here were presented at the 2018 IEEE Conference on Communications and Network Security (CNS) \cite{Huan2018CNS}. Compared with the conference version, the journal version has the following significant differences: 1) the journal version extends the results for constant directed graphs in \cite{Huan2018CNS} to time-varying directed graphs; 2) the journal version provides formal and rigorous analysis of convergence rate that does not exist in the conference version; 3) the journal version allows multiple adversaries to collude to infer the sensitive value a target agent, which is not addressed in our conference version; and 4) the journal version revises and enhances the proposed confidential average consensus algorithm to guarantee information-theoretic privacy, which is stronger than the confidentiality achieved in the conference version.

%%%%%%%%%%%%%%%%%%%%%%%%%%%%%%%%%%%%%%%%%%%%%%%%%%%%%%%%
\section{Preliminaries and Problem Formulation}

%%%%%%%%%%%%%%%%%%%%%%%%%%%%%%%%%%%%%%%%%%%%%%%%%%%%%%%%

\subsection{Graph Representation}
We represent a network of $N$ agents as a sequence of time-varying directed graphs $\{\mathcal{G}(k)=(\mathcal{V}, \, \mathcal{E}(k))\}$ where $\mathcal{V}=\{1, 2, \ldots, N\}$ is the set of agents and $k=0,1,\ldots$ is the time index. $\mathcal{E}(k)\subset \mathcal{V} \times \mathcal{V}$ is the edge set at time $k$, whose elements are such that $(i, \, j) \in \mathcal{E}(k)$ holds if and only if there exists a directed edge from agent $j$ to agent $i$ at time $k$, i.e., agent $j$ can send messages to agent $i$ at time $k$. For notational convenience, we assume that there are no self edges, i.e., $(i, \, i)\notin \mathcal{E}(k)$ for all $k$ and $i\in \mathcal{V}$. At time $k$, each edge $(i, \, j) \in \mathcal{E}(k)$ has an associated weight, $p_{ij}(k)>0$. The out-neighbor set of agent $i$ at time $k$, which represents the set of agents that can receive messages from agent $i$ at time $k$, is denoted as $\mathcal{N}_i^{out}(k)=\{j \in \mathcal{V} \, | \, (j, \, i)\in \mathcal{E}(k)\}$. Similarly, at time $k$, the in-neighbor set of agent $i$, which represents the set of agents that can send messages to agent $i$ at time $k$, is denoted as $\mathcal{N}_i^{in}(k)=\{j \in \mathcal{V} \, | \, (i, \, j)\in \mathcal{E}(k)\}$. From the above definitions, it can be obtained that $i \in \mathcal{N}_j^{out}(k)$ and $j \in \mathcal{N}_i^{in}(k)$ are equivalent. Agent $i$'s out-degree at time instant $k$ is represented by $D_i^{out}(k)=|\mathcal{N}_i^{out}(k)|$ and its in-degree is represented by $D_i^{in}(k)=|\mathcal{N}_i^{in}(k)|$, where $|\mathcal{S}|$ is the cardinality of the set $\mathcal{S}$. 

At iteration $k$, the incidence matrix $\mathbf{C}(k)=[c_{il}(k)]_{N \times E(k)}$ for graph $\mathcal{G}(k)=(\mathcal{V},\mathcal{E}(k))$ is defined as
\begin{equation}
	c_{il}(k)=\left\lbrace \begin{aligned}
		1 \quad & \textnormal{if the} \ l\textnormal{-th edge in} \ \mathcal{E}(k) \ \textnormal{is} \ (i, \, j) \\
		-1 \quad & \textnormal{if the} \ l\textnormal{-th edge in} \ \mathcal{E}(k) \ \textnormal{is} \ (j, \, i) \\
		0 \quad & \textnormal{otherwise}
	\end{aligned} \right. 
\end{equation}
where $E(k)=|\mathcal{E}(k)|$ represents the number of edges in $\mathcal{E}(k)$. Note that the $l$-th column of $\mathbf{C}(k)$ is corresponding to the $l$-th edge in $\mathcal{E}(k)$, and the sum of each column of $\mathbf{C}(k)$ is $0$, i.e., $\mathbf{1}^T \mathbf{C}(k) = \mathbf{0}^T$.

For a sequence of time-varying directed graphs $\{\mathcal{G}(k)=(\mathcal{V}, \, \mathcal{E}(k))\}$, we define $\mathcal{E}_{\infty}$ as the set of directed edges $(i, \, j)$ that exist for infinitely many time instants, i.e.,
\begin{equation}\label{infty_edges_def}
\mathcal{E}_{\infty}=\big\{(i, \, j) \big| (i, \, j) \in \mathcal{E}(k) \ \text{for infinitely many indices} \ k \big\}
\end{equation}

We focus on time-varying directed graphs which satisfy the following assumptions:
\begin{Assumption}\label{assumption_strongly_connected}
	For a sequence of time-varying directed graphs $\{\mathcal{G}(k)=(\mathcal{V}, \, \mathcal{E}(k))\}$, for any $i, j \in \mathcal{V}$ with $i\neq j$, there exists at least one directed path from $i$ to $j$ in $(\mathcal{V}, \, \mathcal{E}_{\infty})$, i.e., $(\mathcal{V}, \, \mathcal{E}_{\infty})$ is strongly connected.
\end{Assumption}
\begin{Assumption}\label{assumption_interval_bound}
	For a sequence of time-varying directed graphs $\{\mathcal{G}(k)=(\mathcal{V}, \, \mathcal{E}(k))\}$, there exists an integer $T \geq 1$ such that for every $(i, \, j) \in \mathcal{E}_{\infty}$, agent $j$ directly communicates with agent $i$ at least once in every $T$ consecutive time instants. $T$ is called intercommunication interval bound.
\end{Assumption}

\begin{Assumption}\label{assumption_out_degree}
	We assume that each agent $i$ has access to its out-degree $D_i^{out}(k)$ at each iteration $k$.
\end{Assumption}

\begin{Remark}\label{remark_out_degree}
	Assumption \ref{assumption_out_degree} is widely used in existing literature on time-varying directed graphs such as \cite{nedic2014distributed, nedic2017achieving, zhu2018differentially}. In fact, in many directed graphs, it is feasible for a node to know its out-neighbors. For example, in many safety-critical systems such as industrial control systems, the exchange of data occurs in a directed way due to unidirectional gateways (aka data diode) whereas control messages (a special type of messages used to configure network connections) can be exchanged in a bidirectional manner to establish connections \cite{scott2015tactical}.
\end{Remark}

%%%%%%%%%%%%%%%%%%%%%%%%%%%%%%%%%%%%%%%%%%%%%%%
\subsection{The Conventional Push-Sum}

The conventional push-sum considers $N$ agents interacting on a constant directed graph $\mathcal{G}=(\mathcal{V}, \, \mathcal{E})$, with each agent having an initial state $x_i^0$ ($i=1,2,\ldots,N$) \cite{kempe2003gossip, benezit2010weighted}. Represent the average value of all initial states as $\bar{x}^0={\sum_{j=1}^{N}x_j^0}/N$. The conventional push-sum algorithm conducts two iterative computations simultaneously, and allows each agent to obtain the exact average of the initial values $\bar{x}^0$ in an asymptotic way. This mechanism is summarized in Algorithm 0 below:

\noindent\rule{0.49\textwidth}{0.5pt}
\noindent\textbf{Algorithm 0: The conventional push-sum algorithm}

\vspace{-0.2cm}\noindent\rule{0.49\textwidth}{0.5pt}
\begin{enumerate}
	\item $N$ agents interact on a constant directed graph $\mathcal{G}=(\mathcal{V}, \, \mathcal{E})$. Each agent $i$ is initialized with $s_i(0)=x_i^0$, $w_i(0)=1$, and $\pi_i(0)=s_i(0)/w_i(0)$. The weight $p_{ij}$ associated with the edge $(i, \, j) \in \mathcal{E}$ satisfies $p_{ij} \in (0,1)$ if $j\in \mathcal{N}_i^{in}\cup \{i\}$ is true and $p_{ij} =0$ otherwise. For any given $j=1, 2, \ldots, N$, $p_{ij}$ satisfies $\sum_{i=1}^N p_{ij}=1$.
	\item At iteration step $k$:
	\begin{enumerate}
		\item Agent $i$ calculates $p_{ji} s_i(k)$ and $p_{ji} w_i(k)$, and sends both values to all of its out-neighbors $j \in \mathcal{N}_i^{out}$.
		\item After receiving the values of $p_{ij} s_j(k)$ and $p_{ij} w_j(k)$ from all its in-neighbors $j \in \mathcal{N}_i^{in}$, agent $i$ updates $s_i$ and $w_i$ as follows:
		\begin{equation}\label{conventional_push_sum}
		\left\lbrace \begin{aligned}
		& \ s_i(k+1) = \sum_{j\in \mathcal{N}_i^{in}\cup \{i\}} p_{ij}s_j(k)\\
		& \ w_i(k+1) = \sum_{j\in \mathcal{N}_i^{in}\cup \{i\}} p_{ij}w_j(k)
		\end{aligned} \right.
		\end{equation}	
		\item Agent $i$ uses the ratio $\pi_i(k+1)=s_i(k+1)/w_i(k+1)$ to estimate the average value $\bar{x}^0={\sum_{j=1}^{N}x_j^0}/N$.
	\end{enumerate}
\end{enumerate}
\vspace{-0.2cm}\rule{0.49\textwidth}{0.5pt}	

For the sake of notational simplicity, we rewrite (\ref{conventional_push_sum}) in the following more compact form:
\begin{equation}\label{conventional_push_sum_vector_form}
\left\lbrace \begin{aligned}
& \ \mathbf{s}(k+1) =\mathbf{P} \mathbf{s}(k)\\
& \ \mathbf{w}(k+1) = \mathbf{P} \mathbf{w}(k)
\end{aligned} \right.
\end{equation}
where $\mathbf{s}(k)= [s_1(k), s_2(k), \ldots, s_N(k)] ^T$ and $\mathbf{w}(k)= [w_1(k), w_2(k), \ldots, w_N(k)]^T$, and $\mathbf{P}=[p_{ij}]$. From Algorithm 0, we have $\mathbf{s}(0)= [x_1^0, x_2^0, \ldots, x_N^0] ^T$ and $\mathbf{w}(0)= \mathbf{1}$. We can also obtain that the matrix $\mathbf{P}$ is column-stochastic, i.e., $\sum_{i=1}^N p_{ij}=1$ holds for $j=1, 2, \ldots, N$.

At iteration step $k$, each agent computes the ratio $\pi_i(k+1)=s_i(k+1)/w_i(k+1)$ to estimate the average value $\bar{x}^0={\sum_{j=1}^{N}x_j^0}/N$. Since $\mathcal{G}$ is assumed to be a strongly connected directed graph, $\mathbf{P}^k$ will converge to a rank-$1$ matrix exponentially fast \cite{Seneta_Markov_book, fill1991eigenvalue}. Defining $\mathbf{P}^\infty$ as the limit of $\mathbf{P}^k$ as $k \rightarrow \infty$, we can obtain the form of $\mathbf{P}^\infty$ as $\mathbf{P}^\infty=\mathbf{v} \mathbf{1}^T$ where $\mathbf{v}=[v_1, v_2, \ldots, v_N]^T$. Using the facts $\mathbf{s}(k)=\mathbf{P}^k \mathbf{s}(0)$ and $\mathbf{w}(k)=\mathbf{P}^k \mathbf{w}(0)$, we can further have \cite{hadjicostis2018distributed}:
\begin{equation}\label{conventional_push_sum_convergence}
\begin{aligned}
\pi_i(\infty)= \frac{s_i(\infty)}{w_i(\infty)}=\frac{ [\mathbf{P}^\infty \mathbf{s}(0)]_i}{[\mathbf{P}^\infty \mathbf{w}(0)]_i}= \frac{v_i \sum_{j=1}^{N} s_j(0)}{v_i \sum_{j=1}^{N} w_j(0)}=\bar{x}^0
\end{aligned}
\end{equation}
where $[\mathbf{P}^\infty \mathbf{s}(0)]_i$ and $[\mathbf{P}^\infty \mathbf{w}(0)]_i$ represent the $i$-th element of vector $\mathbf{P}^\infty \mathbf{s}(0)$ and vector $\mathbf{P}^\infty \mathbf{w}(0)$, respectively. Hence, all estimates $\pi_1(k), \pi_2(k), \ldots, \pi_N(k)$ will asymptotically converge to the average $\bar{x}^0={\sum_{j=1}^{N}x_j^0}/N$.

%%%%%%%%%%%%%%%%%%%%%%%%%%%%%%%%%%%%%%%%%%%%%%
\subsection{Problem Formulation}

In this paper, we will address average consensus under time-varying directed graphs while protecting the confidentiality of participating agents against adversaries. To this end, we first present some assumptions and definitions.

\begin{Assumption}\label{assumption_initial_states}
	We assume that all agents' initial states $x_i^0$ are bounded. Without loss of generality, the lower bound and upper bound are denoted as $a$ and $b$, respectively. Both $a$ and $b$ are assumed known to all agents.
\end{Assumption}

\begin{Remark}\label{remark_initial_states}
	It is worth noting that although the bounds $a$ and $b$ are assumed known to all agents, this does not mean that the minimum and maximum of all agents' initial states are known to all agents. In fact, $a$ (resp. $b$) can be arbitrarily small (resp. large) compared with the actual minimum (resp. maximum) of agents' initial states.
\end{Remark}

\begin{Definition}\label{honest_but_curious_definition}
	We define an honest-but-curious adversary as an agent who follows all protocol steps correctly but collects received messages in an attempt to infer the initial value of other participating agents.
\end{Definition}

\begin{Assumption}\label{assumption_collude}
	We assume that agents can collude, i.e., a set of honest-but-curious agents $\mathcal{A}$ can share information with each other to infer the initial value $x_i^0$ of a target agent $i \notin \mathcal{A}$.
\end{Assumption}

\begin{Definition}\label{privacy_preservation_definition}
	We define that confidentiality (privacy) of the initial value $x_i^0$ of agent $i$ is preserved if $x_i^0$ is indistinguishable from the viewpoint of honest-but-curious adversaries $\mathcal{A}$. By ``indistinguishable,'' we mean that the probability distribution of information set accessible to $\mathcal{A}$ does not change when agent $i$'s initial state $x_i^0$ is altered to any $\tilde{x}_i^0 \neq x_i^0$ under the constraint that the sum of the initial states of all nodes not in $\mathcal{A}$ (i.e., $\sum_{ j\in \mathcal{V}\setminus\mathcal{A}} x_j^0$) is unchanged.
\end{Definition}

Our definition of confidentiality requires perfect indistinguishability of a target agent's different initial states from the viewpoint of honest-but-curious adversaries $\mathcal{A}$, and, therefore, is more stringent than the confidentiality definition in \cite{manitara2013privacy, duan2015privacy, liu2006random, han2010privacy, cao2014privacy} which defines confidentiality as the inability of an adversary to {\it uniquely} determine the sensitive value.

We next show that the conventional push-sum is not confidential. From (\ref{conventional_push_sum}) and (\ref{conventional_push_sum_vector_form}), an honest-but-curious agent $i$ can receive $p_{ij}s_j(0)$ and $p_{ij}w_j(0)$ from its in-neighbor agent $j$ after the first iteration step $k=0$. Then agent $i$ is able to uniquely determine $x_j^0$ by $x_j^0 =s_j(0) = \frac{p_{ij}s_j(0)} {p_{ij}w_j(0)}$ using the fact $w_j(0)=1$. Therefore, an honest-but-curious agent can always infer the initial values of all its in-neighbors, and hence the conventional push-sum algorithm cannot provide confidentiality against honest-but-curious adversaries. It is worth noting that using a similar argument, we can also obtain that the conventional push-sum is not confidential even when the weight is allowed to be time-varying (e.g., \cite{benezit2010weighted}.)

%%%%%%%%%%%%%%%%%%%%%%%%%%%%%%%%%%%%%%%%%%%%%%%%%%%%%%%%%%%%%%%%%%%%%%%%%%%%%%%%%%%%%%%%%%%%%%%%%%%%%%%%%%%%%%%%%%%%%%%
\section{The Confidentiality Algorithm and Performance Analysis}

In this section, we will propose a confidential average consensus algorithm for time-varying directed graphs, and then provide rigorous analysis of its convergence rate and enabled strength of confidentiality.

%%%%%%%%%%%%%%%%%%%%%%%%%%%%%%%%%%%%%%%%%%%%%%%%%%%%%%%%%%%%%%%%%%%%%%%%
\subsection{Confidential Average Consensus Algorithm}

The analysis above reveals that using the same weight $p_{ij}$ for both $p_{ij}s_j(0)$ and $p_{ij}w_j(0)$ discloses the initial state value. Motivated by this observation and the work in \cite{gupta2019statistical}, here we introduce a novel confidential average consensus algorithm which injects randomness in the dynamics of interactions in iterations $k=0, \ldots, K$. Note that here $K$ is a non-negative integer and is known to every agent. Its influence will be discussed in detail in Remark \ref{remark_tradeoff_privacy_convergence} and Remark \ref{remark_tradeoff_intermediate_privacy}.

\noindent\rule{0.49\textwidth}{0.5pt}
\noindent\textbf{Algorithm 1: Confidential average consensus algorithm}

\vspace{-0.2cm}\noindent\rule{0.49\textwidth}{0.5pt}
\begin{enumerate}
	\item $N$ agents interact on a sequence of time-varying directed graphs $\{\mathcal{G}(k)=(\mathcal{V}, \, \mathcal{E}(k))\}$. Each agent $i$ is initialized with $s_i(0)= \frac{1}{N^2} + \frac{(N-2)(x_i^0-a)}{(b-a)N^2} \in [\frac{1}{N^2}, \, \frac{N-1}{N^2}]$, $w_i(0)=1$, and $\pi_i(0)=\frac{b-a}{N-2}[N \times {\rm frac}(\frac{Ns_i(0)}{w_i(0)})-1]+a$ where the function ${\rm frac}(x)=x-\lfloor x \rfloor$ denotes the fractional part of a real number $x$ (here $\lfloor x \rfloor$ represents the largest integer not greater than $x$).
	\item At iteration step $k$:
	\begin{enumerate}
		\item Agent $i$ generates a set of random weights $\big\{p_{ji}(k) \in (\varepsilon, \, 1) \, \big| \, j\in \mathcal{N}_i^{out}(k) \cup \{i\}\big\}$ with the sum of this set equal to $1$, and sets $\Delta w_{ji}(k) = p_{ji}(k) w_i(k)$ for $j\in \mathcal{N}_i^{out}(k) \cup \{i\}$.
		\item If $k \leq K$, agent $i$ independently generates uniformly distributed values $\Delta s_{ji}(k) \in [0, \, 1)$ for its out-neighbors $j\in \mathcal{N}_i^{out}(k)$, and sets $\Delta s_{ii}(k)={\rm frac}\big(s_i(k)-\sum_{j\in \mathcal{N}_i^{out}(k)}\Delta s_{ji}(k)\big)$; otherwise, agent $i$ sets $\Delta s_{ji}(k) = p_{ji}(k) s_i(k)$ for $j\in \mathcal{N}_i^{out}(k) \cup \{i\}$.
		\item Agent $i$ sends $\Delta s_{ji}(k)$ and $\Delta w_{ji}(k)$ to its out-neighbors $j\in \mathcal{N}_i^{out}(k)$.
		\item After receiving $\Delta s_{ij}(k)$ and $\Delta w_{ij}(k)$ from its in-neighbors $j \in \mathcal{N}_i^{in}(k)$, agent $i$ updates $s_i$ and $w_i$ as
		\begin{center}
			\begin{center}
				\begin{equation}\label{Algorithm_I_s_update} 
			s_i(k+1) = \left\lbrace \begin{aligned}
				&  {\rm frac}\Big(\sum_{j\in \mathcal{N}_i^{in}(k) \cup \{i\}} \Delta s_{ij}(k)\Big) \ \textnormal{for} \, k\leq K \\
				& \sum_{j\in \mathcal{N}_i^{in}(k) \cup \{i\}} \Delta s_{ij}(k) \quad \ \textnormal{for} \, k \geq K+1 \\
			\end{aligned} \right.
		\end{equation}
			\end{center}
		\end{center}
		and
		\begin{equation}\label{Algorithm_I_w_update}
			\begin{aligned}
				w_i(k+1) = \sum_{j\in \mathcal{N}_i^{in}(k) \cup \{i\} } \Delta w_{ij}(k) \quad \textnormal{for} \ k \geq 0\\
			\end{aligned}
		\end{equation}
		respectively.	
		\item Agent $i$ uses the ratio $\pi_i(k+1)=\frac{b-a}{N-2}[N \times {\rm frac}(\frac{Ns_i(k+1)}{w_i(k+1)})-1]+a$ to estimate the average value $\bar{x}^0={\sum_{j=1}^{N}x_j^0}/N$.
	\end{enumerate}
\end{enumerate}
\vspace{-0.2cm}\rule{0.49\textwidth}{0.5pt}

\begin{Remark}
	Compared to the conventional confidentiality-violating push-sum algorithm which broadcasts messages, Algorithm 1 needs agent $i$ to send different random numbers to different out-neighbors in iterations $k \leq K$. This is a price of obtaining confidentiality without losing accuracy in the time-varying directed topology case.
\end{Remark}

\begin{Remark}
	The way of injecting randomness in $\Delta s_{ji}(k)$ is different in iterations $k\leq K$ from $k>K$. In fact, in iterations $k\leq K$, $\Delta s_{ji}(k)$ can be nonzero even when $s_i(k)$ is zero. This is crucial in enabling strong confidentiality as receiving $\Delta s_{ji}(k)$ of a value zero will not allow the recipient to infer information about $s_i(k)$.
\end{Remark}

\begin{Remark}\label{remark_random_weights}
	 In Algorithm 1, the coupling weights are randomly chosen at every iteration. Compared with the commonly used deterministic setting in which the weights are set as $p_{ji}(k)=1/(D_i^{out}(k)+1)$ for all $j \in \mathcal{N}_i^{out}(k) \cup \{i\}$, our setting is more general since it includes the commonly used setting as a special case by fixing $p_{ji}(k)$ to deterministic values. Furthermore, the random weights beyond step $K$ in Algorithm 1 can provide additional confidentiality protection for intermediate states $s_i(k)$ after iteration $K$. Given $\Delta s_{ji}(k)= p_{ji}(k) s_i(k)$ for $k \geq K+1$, we can see that using random weights $p_{ji}(k)$ makes the intermediate state $s_i(k)$ more difficult to infer than using deterministic weights $p_{ji}(k)$.
\end{Remark}

Setting $\Delta s_{ji}(k)$, $\Delta w_{ji}(k)$, and $p_{ji}(k)$ to $0$ for $j\notin \mathcal{N}_i^{out}(k)\cup \{i\}$, we can rewrite the update rules of $\mathbf{s}(k)$ for $k\geq K+1$ and $\mathbf{w}(k)$ for $k\geq 0$ as
\begin{equation}\label{Algorithm_s_w}
	\left\lbrace \begin{aligned}
		& s_i(k+1) = \sum_{j=1}^N \Delta s_{ij}(k) = \sum_{j=1}^N p_{ij}(k)s_j(k) \ \, \textnormal{for} \ k \geq K+1 \\
		& w_i(k+1) = \sum_{j=1}^N \Delta w_{ij}(k)=\sum_{j=1}^N p_{ij}(k)w_j(k) \ \ \textnormal{for} \ k \geq 0
	\end{aligned} \right.
\end{equation}
Denoting $\mathbf{s}(k)$, $\mathbf{w}(k)$, and $\mathbf{P}(k)$ as $\mathbf{s}(k)= [s_1(k)\, \cdots \, s_N(k)] ^T$, $\mathbf{w}(k)= [w_1(k) \, \cdots \, w_N(k)]^T$, and $\mathbf{P}(k)=[p_{ij}(k)]_{N\times N}$, we can further rewrite (\ref{Algorithm_s_w}) into a matrix form
\begin{equation}\label{our_algorithm_I}
\left\lbrace \begin{aligned}
& \ \mathbf{s}(k+1) =\mathbf{P}(k) \mathbf{s}(k) \quad \ \ \text{for} \ k \geq K+1 \\
& \ \mathbf{w}(k+1) = \mathbf{P}(k) \mathbf{w}(k) \quad \text{for} \ k \geq 0
\end{aligned} \right.
\end{equation}
For iteration $k=0$, we have $\mathbf{s}(0)= [x_1^0, x_2^0, \ldots, x_N^0] ^T$ and $\mathbf{w}(0)= \mathbf{1}$. From Algorithm 1, we know that $\mathbf{P}(k)$ in (\ref{our_algorithm_I}) is time-varying and column-stochastic for $k \geq 0$.

Defining the transition matrix as follows
\begin{equation}\label{transition_matrix}
\begin{aligned}
\mathbf{\Phi}(k:t) =\mathbf{P}(k) \cdots\mathbf{P}(t)\\
\end{aligned}
\end{equation}
for all $k$ and $t$ with $k \geq t$, where $\mathbf{\Phi}(k:k)=\mathbf{P}(k)$, we can rewrite (\ref{our_algorithm_I}) as
\begin{equation}\label{Algorithm_I_second_half}
\left\lbrace \begin{aligned}
& \mathbf{s}(k+1) = \mathbf{\Phi}(k:K+1) \mathbf{s}(K+1) \quad \text{for} \ k \geq K+1 \\
& \mathbf{w}(k+1) = \mathbf{\Phi}(k:0) \mathbf{w}(0) \qquad \text{for} \ k \geq 0
\end{aligned} \right.
\end{equation}

%%%%%%%%%%%%%%%%%%%%%%%%%%%%%%%%%%%%%%%%%%%%%%%%%%%%%%%%%%%%%%%%%%%%%%%%
\subsection{Convergence Analysis}

Next we prove that Algorithm 1 can guarantee that the estimates of all agents converge to the exact average value of initial values. We will also analyze the rate of convergence of Algorithm 1. Using the convergence definition in \cite{nedic2017achieving} and \cite{nedich2016geometrically}, we define the rate of convergence to be at least $\gamma \in (0, \, 1)$ if there exists a positive constant value $C$ such that $\big\|\boldsymbol{\pi}(k) - \bar{x}^0 \mathbf{1} \big\| \leq C \gamma ^k$ is true for all $k$, where $\boldsymbol{\pi}(k)=[\pi_1(k), \ldots, \pi_N(k)]^{T}$ and $\bar{x}^0={\sum_{j=1}^{N}x_j^0}/N$ is the average value. Note that this definition means a smaller $\gamma$ corresponding to a faster convergence. To analyze the convergence rate of Algorithm 1, we first introduce Lemma \ref{lemma_1} below:

\begin{Lemma}\label{lemma_1}
	For a network of $N$ agents represented by a sequence of time-varying directed graphs $\{\mathcal{G}(k)=(\mathcal{V}, \, \mathcal{E}(k))\}$ which satisfy Assumptions \ref{assumption_strongly_connected}, \ref{assumption_interval_bound}, and \ref{assumption_out_degree}, under Algorithm 1, each agent $i$ has $w_i(k) \geq \varepsilon^{T(N-1)}$ for $k \geq 1$ where $T$ is defined in Assumption \ref{assumption_interval_bound}.
\end{Lemma}

{\it Proof}: For $k \geq 1$, from (\ref{Algorithm_I_second_half}) we have
\begin{equation}\label{Lemma_1_01} 
\begin{aligned}
\mathbf{w}(k) = \mathbf{\Phi}(k-1:0) \mathbf{1}
\end{aligned}
\end{equation}	
Represent $\delta(k)$ as
\begin{equation}\label{Lemma_1_02}
\begin{aligned}
\delta(k) \triangleq \min_{1 \leq i \leq N}{w}_i(k) = \min_{1 \leq i \leq N} [\mathbf{\Phi}(k-1:0) \, \mathbf{1}]_i
\end{aligned}
\end{equation}
for $k \geq 1$. To prove $w_i(k) \geq \varepsilon^{T(N-1)}$ for $k \geq 1$, it is sufficient to prove $\delta(k) \geq \varepsilon^{T(N-1)}$ for $k \geq 1$. We divide our proof into two parts: $1 \leq k \leq T(N-1)$ and $k \geq T(N-1)+1$.

\textbf{Part 1:} $\delta(k) \geq \varepsilon^{T(N-1)}$ for $1 \leq k \leq T(N-1)$. One can verify that the following relationship holds
\begin{equation} 
\begin{aligned}
& [\mathbf{\Phi}(k-1:0)]_{ii} \\
= \ \ & [\mathbf{P}(k-1) \cdots \mathbf{P}(0)]_{ii} \\
\geq \ \ & [\mathbf{P}(k-1)]_{ii} \, [\mathbf{P}(k-2) \cdots \mathbf{P}(0)]_{ii}\\
\geq \ \ & \varepsilon \, [\mathbf{\Phi}(k-2:0)]_{ii}
\end{aligned} 
\end{equation}
Given $[\mathbf{\Phi}(0:0)]_{ii}=[\mathbf{P}(0)]_{ii} \geq \varepsilon$, one can obtain $[\mathbf{\Phi}(k-1:0)]_{ii} \geq \varepsilon^k$. Therefore, it follows that
\begin{equation} 
\begin{aligned}
\left[\mathbf{\Phi}(k-1:0) \, \mathbf{1}\right]_i & \geq [\mathbf{\Phi}(k-1:0)]_{ii} \\
& \geq \varepsilon^k \geq \varepsilon^{T(N-1)}
\end{aligned}
\end{equation}	
is true for $i=1,\ldots, N$ and $1 \leq k \leq T(N-1)$, implying $\delta(k) \geq \varepsilon^{T(N-1)}$ for $1 \leq k \leq T(N-1)$.

\textbf{Part 2:} $\delta(k) \geq \varepsilon^{T(N-1)}$ for $k \geq T(N-1)+1$. Under Assumptions \ref{assumption_strongly_connected} and \ref{assumption_interval_bound}, and the requirements on weights $p_{ij}(k)$ in Algorithm 1, and following the arguments in Lemma 2 in \cite{nedic2009distributed}, we can obtain
\begin{equation}
[\mathbf{\Phi}(k-1:k-T(N-1))]_{ij} \geq \varepsilon^{T(N-1)}
\end{equation}
for $1 \leq i, \, j \leq N$. Since $k\geq T(N-1)+1$ holds and $\mathbf{P}(k)$ is a column-stochastic matrix,
\begin{equation}
\mathbf{\Phi}(k-T(N-1)-1:0) = \mathbf{P}(k-T(N-1)-1) \cdots \mathbf{P}(0)
\end{equation}
should also be a column-stochastic matrix. Further using the fact
$\mathbf{\Phi}(k-1:0) = \mathbf{\Phi}(k-1:k-T(N-1)) \mathbf{\Phi}(k-T(N-1)-1:0)$ leads to $[\mathbf{\Phi}(k-1:0)]_{ij} \geq \varepsilon^{T(N-1)}$ for $1 \leq i, \, j \leq N$. Therefore, we have
\begin{equation}
[\mathbf{\Phi}(k-1:0) \, \mathbf{1}]_i \geq N \varepsilon^{T(N-1)} \geq \varepsilon^{T(N-1)}
\end{equation}
for $i=1,\ldots, N$, meaning $\delta(k) \geq \varepsilon^{T(N-1)}$ for $k\geq T(N-1)+1$.

Based on $\delta(k) \geq \varepsilon^{T(N-1)}$ for $k \geq 1$, we can obtain $w_i(k) \geq \varepsilon^{T(N-1)}$ for $k \geq 1$. In summary, we always have $w_i(k) \geq \varepsilon^{T(N-1)}$ for $k \geq 1$. \hfill{$\blacksquare$}

\begin{Theorem}\label{theorem_convergence}
	For a network of $N$ agents represented by a sequence of time-varying directed graphs $\{\mathcal{G}(k)=(\mathcal{V}, \, \mathcal{E}(k))\}$ satisfying Assumptions \ref{assumption_strongly_connected}, \ref{assumption_interval_bound}, \ref{assumption_out_degree}, and \ref{assumption_initial_states}, under Algorithm 1, the estimate $\pi_i(k)=\frac{b-a}{N-2}[N \times {\rm frac}(\frac{Ns_i(k)}{w_i(k)})-1]+a$ of each agent $i$ will converge to the average $\bar{x}^0={\sum_{j=1}^{N}x_j^0}/N$. More specifically, the rate of convergence of Algorithm 1 is at least $\gamma= (1-\varepsilon^{T(N-1)}) ^{\frac{1}{T(N-1)}} \in (0, \, 1)$, meaning that there exists a positive constant value $C$ satisfying $\big\|\boldsymbol{\pi}(k) - \bar{x}^0 \mathbf{1} \big\| \leq C \gamma ^k$ for all $k$.
\end{Theorem}

{\it Proof}: From (\ref{Algorithm_I_s_update}), we have
\begin{equation}\label{mass_conservation_property}
	\begin{aligned}
		& {\rm frac}\Big(\sum_{i=1}^{N}s_i(k+1)\Big)\\ 
		= & {\rm frac}\bigg(\sum_{i=1}^{N} {\rm frac}\Big(\sum_{j \in \mathcal{N}_i^{in}(k)\cup \{i\}} \Delta s_{ij}(k) \Big)\bigg)\\
		= & {\rm frac}\bigg(\sum_{i=1}^{N} \sum_{j \in \mathcal{N}_i^{in}(k)} \Delta s_{ij}(k) + \sum_{i=1}^{N} \Delta s_{ii}(k) \bigg)\\
		= & {\rm frac}\bigg(\sum_{i=1}^{N} \sum_{j \in \mathcal{N}_i^{in}(k)} \Delta s_{ij}(k) \\
		& \qquad \quad + \sum_{i=1}^{N} {\rm frac}\Big(s_i(k)-\sum_{j\in \mathcal{N}_i^{out}(k)}\Delta s_{ji}(k)\Big) \bigg)\\
		= & {\rm frac}\Big(\sum_{i=1}^{N}s_i(k)\Big)
	\end{aligned}
\end{equation}
for $k \leq K$ where in the derivation we used the property ${\rm frac}(x+y) = {\rm frac}(x+{\rm frac}(y)) = {\rm frac}({\rm frac}(x)+y) = {\rm frac}({\rm frac}(x)+{\rm frac}(y))$ for any $x, \, y \in \mathbb{R}$. According to Assumption \ref{assumption_initial_states}, $x_i^0 \in [a, \, b]$ holds for each agent $i$. Given $s_i(0)= \frac{1}{N^2} + \frac{(N-2)(x_i^0-a)}{(b-a)N^2}$, one can obtain $s_i(0)\in [\frac{1}{N^2}, \, \frac{N-1}{N^2}]$, leading to
\begin{equation}\label{eqn_initial_sum}
	{\rm frac}\big(\sum_{i=1}^{N}s_i(0)\big) = \sum_{i=1}^{N}s_i(0) \in [\frac{1}{N}, \, \frac{N-1}{N}] \subset (0, \, 1)
\end{equation}
Further combining (\ref{mass_conservation_property}) and (\ref{eqn_initial_sum}) yields
\begin{equation}\label{eqn_state_sum}
	{\rm frac}\big(\sum_{i=1}^{N}s_i(K+1)\big) = \cdots = {\rm frac}\big(\sum_{i=1}^{N}s_i(0)\big) = \sum_{i=1}^{N}s_i(0)
\end{equation}
Since $\mathbf{P}(k)$ is column stochastic, from (\ref{our_algorithm_I}) we have
\begin{equation}\label{mass_conservation_w}
	\begin{aligned}
		\mathbf{1}^T \mathbf{w}(k+1) = \mathbf{1}^T \mathbf{w}(k) = \cdots =\mathbf{1}^T \mathbf{w}(0) = N
	\end{aligned}
\end{equation}
for $k \geq 0$. Then we rewrite (\ref{Algorithm_I_second_half}) as
\begin{equation}\label{Theorem_1_second_half}
\left\lbrace \begin{aligned}
& \mathbf{s}(K+l+1) = \mathbf{\Phi}(K+l:K+1) \mathbf{s}(K+1)\\
& \mathbf{w}(K+l+1) = \mathbf{\Phi}(K+l:K+1) \mathbf{w}(K+1)\\
\end{aligned} \right.
\end{equation}
for $l \geq 1$. Under Assumptions \ref{assumption_strongly_connected} and \ref{assumption_interval_bound}, and the requirements on weights $p_{ij}(k)$ in Algorithm 1, following Proposition 1(b) in \cite{nedic2010constrained}, we know that the transition matrix $\mathbf{\Phi}(K+l:K+1)$ will converge to a stochastic vector $\boldsymbol{\varphi}(K+l)$ with a geometric rate for all $i$ and $j$, i.e., for all $i, j=1,\ldots,N$ and $l \geq 1$, we have
\begin{equation}\label{Phi_i_j}
\begin{aligned}
\big|[\mathbf{\Phi}(K+l:K+1)]_{ij}-\varphi_i(K+l)\big| \leq C_0 \gamma^{l-1}
\end{aligned}
\end{equation}
with $C_0=2 ({1+\varepsilon^{-T(N-1)}})/({1-\varepsilon^{T(N-1)}})$ and $\gamma= (1-\varepsilon^{T(N-1)}) ^{\frac{1}{T(N-1)}}$. Defining $\mathbf{M}(K+l:K+1)$ as
\begin{equation}\label{D_matrix}
\begin{aligned}
\mathbf{M}(K+l:K+1) \triangleq \mathbf{\Phi}(K+l:K+1)-\boldsymbol{\varphi}(K+l) \, \mathbf{1}^T
\end{aligned}
\end{equation}
we have
\begin{equation}\label{D_i_j}
\begin{aligned}
\left|\left[\mathbf{M}(K+l:K+1)\right]_{ij}\right| \leq C_0 \gamma^{l-1}
\end{aligned}
\end{equation}
for all $i, j=1,\ldots,N$ and $l \geq 1$. Further combining (\ref{D_matrix}) with (\ref{Theorem_1_second_half}) leads to
\begin{equation}\label{Theorem_1_second_half_new}
\left\lbrace \begin{aligned}
\mathbf{s}(K+l+1)= & \mathbf{M}(K+l:K+1) \mathbf{s}(K+1) \\
& + \, \boldsymbol{\varphi}(K+l) \, \mathbf{1}^T \mathbf{s}(K+1) \\
\mathbf{w}(K+l+1)= & \mathbf{M}(K+l:K+1) \mathbf{w}(K+1) \\
& + \, N \boldsymbol{\varphi}(K+l)\\
\end{aligned} \right.
\end{equation}
where in the derivation we used $\mathbf{1}^T \mathbf{w}(K+1)=N$ from (\ref{mass_conservation_w}). Then from (\ref{Theorem_1_second_half_new}), we have
\begin{equation}\label{pi_alpha}
\begin{aligned}
& \frac{s_i(K+l+1)}{w_i(K+l+1)}- \frac{\sum_{j=1}^{N}s_j(K+1)}{N} \\
= \, & \frac{s_i(K+l+1)}{w_i(K+l+1)}-\frac{\mathbf{1}^T \mathbf{s}(K+1)}{N} \\
= \, & \frac{s_i(K+l+1)}{w_i(K+l+1)} - \frac{\mathbf{1}^T \mathbf{s}(K+1) w_i(K+l+1)}{N w_i(K+l+1)}\\
= \, & \frac{[\mathbf{M}(K+l:K+1) \mathbf{s}(K+1)]_i + \varphi_i(K+l) \mathbf{1}^T \mathbf{s}(K+1)} {w_i(K+l+1)}\\
& \, - \frac{\mathbf{1}^T \mathbf{s}(K+1) [\mathbf{M}(K+l:K+1) \mathbf{w}(K+1)]_i}{N w_i(K+l+1)}\\
& \, - \frac{\mathbf{1}^T \mathbf{s}(K+1) N \varphi_i(K+l)}{N w_i(K+l+1)}\\
= \, & \frac{[\mathbf{M}(K+l:K+1) \mathbf{s}(K+1)]_i} {w_i(K+l+1)} \\
& \, - \frac{\mathbf{1}^T \mathbf{s}(K+1) [\mathbf{M}(K+l:K+1) \mathbf{w}(K+1)]_i} {N w_i(K+l+1)}\\
\end{aligned}
\end{equation}
Therefore, for $i=1,\ldots,N$ and $l \geq 1$, we can obtain
\begin{equation}\label{pi_alpha_absolute}
\begin{aligned}
& \big| N \frac{s_i(K+l+1)}{w_i(K+l+1)}- \sum_{j=1}^{N}s_j(K+1) \big| \\
\leq & \frac{N\big| [\mathbf{M}(K+l:K+1) \mathbf{s}(K+1)]_i \big|} {w_i(K+l+1)} \\
& \, + \frac{N\big| \mathbf{1}^T \mathbf{s}(K+1) [\mathbf{M}(K+l:K+1) \mathbf{w}(K+1)]_i \big|} {N w_i(K+l+1)}\\
\leq & \frac{N}{\varepsilon^{T(N-1)}} \big(\max_{j}\big|[\mathbf{M}(K+l:K+1)]_{ij} \big|\big) \big\|\mathbf{s}(K+1)\big\|_1 \\
& \, + \frac{N}{\varepsilon^{T(N-1)}} \big|\mathbf{1}^T \mathbf{s}(K+1)\big| \big(\max_{j}\big|[\mathbf{M}(K+l:K+1)]_{ij} \big|\big)\\
\end{aligned}
\end{equation}
where in the derivation we used $w_i(K+l+1) \geq \varepsilon^{T(N-1)}$ from Lemma \ref{lemma_1} and $\big\|\mathbf{w}(K+1)\big\|_1= \sum_{i=1}^{N} |w_i(K+1)|= \mathbf{1}^T \mathbf{w}(K+1) =N$ from (\ref{mass_conservation_w}). Further using the relationship $\big|\mathbf{1}^T \mathbf{s}(K+1)\big| \leq \big\| \mathbf{s}(K+1) \big\|_1$ and (\ref{D_i_j}) yields
\begin{equation}\label{pi_alpha_absolute_new}
	\begin{aligned}
		\big| N \frac{s_i(k)}{w_i(k)}- \sum_{j=1}^{N}s_j(K+1) \big| \leq C_1 \gamma^k
	\end{aligned}
\end{equation}
for $k\geq K+2$ with $C_1$ given by
\begin{equation}\label{eqn_C_1}
	\begin{aligned}
		C_1=2 N C_0 \big\|\mathbf{s}(K+1)\big\|_1 \varepsilon^{-T(N-1)} \gamma^{-K-2}
	\end{aligned}
\end{equation}

Given $s_i(0)= \frac{1}{N^2} + \frac{(N-2)(x_i^0-a)}{(b-a)N^2}$, we have
\begin{equation}\label{eqn_initial_x_s}
	\begin{aligned}
		\bar{x}^0 & = \frac{1}{N} \sum_{j=1}^{N}x_j^0 = \frac{b-a}{N-2}\Big(N\sum_{j=1}^{N}s_j(0)-1\Big)+a \\
		& = \frac{b-a}{N-2}\Big(N \times {\rm frac}\big(\sum_{j=1}^{N} s_j(K+1)\big)-1\Big)+a
	\end{aligned}
\end{equation}
where in the derivation we used ${\rm frac}\big(\sum_{j=1}^{N} s_j(K+1)\big) = \sum_{j=1}^{N} s_j(0)$ from (\ref{eqn_state_sum}). Combining (\ref{eqn_initial_x_s}) with $\pi_i(k)=\frac{b-a}{N-2}[N \times {\rm frac}(\frac{Ns_i(k)}{w_i(k)})-1]+a$ leads to
\begin{equation}\label{eqn_estimate_error}
	\begin{aligned}
		& \pi_i(k)-\bar{x}^0 \\
		= \, & \frac{b-a}{N-2}N\Big({\rm frac}\big( \frac{Ns_i(k)}{w_i(k)} \big) - {\rm frac}\big( \sum_{j=1}^{N} s_j(K+1) \big) \Big)
	\end{aligned}
\end{equation}
From (\ref{eqn_initial_sum}) and (\ref{eqn_state_sum}), one can obtain ${\rm frac}\big( \sum_{j=1}^{N} s_j(K+1) \big)=\sum_{j=1}^{N} s_j(0) \in [\frac{1}{N}, \, \frac{N-1}{N}]$. Defining $\eta$ as $\eta \triangleq \sum_{j=1}^{N} s_j(0) \in [\frac{1}{N}, \, \frac{N-1}{N}]$, then there must exist an integer $Q$ such that $\sum_{j=1}^{N} s_j(K+1)=Q+\eta$ holds. From (\ref{pi_alpha_absolute_new}), we can see that there must exist a positive integer $K_1 \geq K+2$ such that
\begin{equation}\label{eqn_estimate_error_1}
	\begin{aligned}
		\big| N \frac{s_i(k)}{w_i(k)}- \sum_{j=1}^{N}s_j(K+1) \big| \leq C_1 \gamma^k < \frac{1}{N}
	\end{aligned}
\end{equation}
holds for $k \geq K_1$. Then it follows naturally one has
\begin{equation}
	\begin{aligned}
		Q<Q+\eta-C_1 \gamma^k \leq N \frac{s_i(k)}{w_i(k)} \leq Q+\eta+C_1 \gamma^k < Q+1
	\end{aligned}
\end{equation}
for $k \geq K_1$, which leads to
\begin{equation}
	\begin{aligned}
		0< \eta-C_1 \gamma^k \leq {\rm frac}\big(N \frac{s_i(k)}{w_i(k)}\big) \leq \eta+C_1 \gamma^k < 1
	\end{aligned}
\end{equation}
for $k \geq K_1$. Thus, we have
\begin{equation}\label{eqn_estimate_error_2}
	\begin{aligned}
		\big| \pi_i(k)-\bar{x}^0 \big| 
		& = \frac{b-a}{N-2}N\big|{\rm frac}\big(\frac{Ns_i(k)}{w_i(k)} \big) - \eta \big| \\
		& \leq \frac{b-a}{N-2}N C_1 \gamma^k
	\end{aligned}
\end{equation}
and further
\begin{equation}\label{eqn_estimate_error_3}
	\begin{aligned}
		\big\| \boldsymbol{\pi}(k)-\bar{x}^0\mathbf{1} \big\| \leq C_2 \gamma^k
	\end{aligned}
\end{equation}
for $k \geq K_1$ with $C_2 \triangleq \frac{b-a}{N-2}N \sqrt{N} C_1$.

For $k \leq K_1-1$, from (\ref{eqn_estimate_error}), we have
\begin{equation}
	\begin{aligned}
		|\pi_i(k)-\bar{x}^0| < \frac{b-a}{N-2}N
	\end{aligned}
\end{equation}
which further implies
\begin{equation}
	\begin{aligned}
		\|\boldsymbol{\pi}(k)-\bar{x}^0\mathbf{1} \| < \frac{b-a}{N-2}N\sqrt{N}
	\end{aligned}
\end{equation}
for $k \leq K_1-1$. Defining $C$ as
\begin{equation}\label{C}
	\begin{aligned}
		C \triangleq \max \big\{C_2, \frac{b-a}{N-2}N \sqrt{N} \gamma^{-k} \, \big| \, 0 \leq k \leq K_1-1 \big\}
	\end{aligned}
\end{equation}
one can obtain
\begin{equation}\label{pi_alpha__vector_all}
	\begin{aligned}
		\big\|\boldsymbol{\pi}(k) - \bar{x}^0 \mathbf{1} \big\| \leq C \gamma^{k}
	\end{aligned}
\end{equation}
for all $k$. Therefore, each agent $i$ will converge to the average value $\bar{x}^0={\sum_{j=1}^{N}x_j^0}/N$ with the rate of convergence of at least $\gamma= (1-\varepsilon^{T(N-1)})^{\frac{1}{T(N-1)}} \in (0, \, 1)$. \hfill{$\blacksquare$}

From Theorem \ref{theorem_convergence}, we can see that a smaller $\gamma$ means a faster convergence. Under the relationship $\gamma= (1-\varepsilon^{T(N-1)}) ^{\frac{1}{T(N-1)}}$, to expedite the convergence, i.e., a smaller $\gamma$, it is sufficient to increase $\varepsilon$, which amounts to reducing the width of the range $(\varepsilon, \, 1)$ for the random selection of $p_{ji}(k)$. Note that although a reduced range $(\varepsilon, \, 1)$ enables an honest-but-curious adversary to obtain a better estimation of the range of agent $i$'s intermediate states $s_i(k)$ and $w_i(k)$ for $k \geq K+1$ from received $p_{ji}(k)s_i(k)$ and $p_{ji}(k)w_i(k)$, it does not affect the confidentiality of agent $i$'s initial state $x_i^0$, as will be shown in the following subsection. It is also worth noting that to meet the requirement of randomly selecting weights in our algorithm, $\varepsilon$ cannot be arbitrarily close to $1$. In fact, $\varepsilon$ must satisfy $\varepsilon < 1/\max_{i,k}({D_i^{out}(k)}+1)$. An easy way to select $\varepsilon$ is to set $0<\varepsilon<1/N$ since $D_i^{out}(k) \leq N-1$ is true for all $k$ and $i \in \mathcal{V}$.

\begin{Remark}\label{remark_time_varying_convergence}
	Theorem \ref{theorem_convergence} provides a detailed analysis of the rate of convergence under time-varying directed graphs, the results on which are sparse in the literature on Push-Sum under time-varying random weights.
\end{Remark}

\subsection{Confidentiality Analysis}

Before presenting our main confidentiality result, we first introduce the following lemma.

\begin{Lemma}\label{lemma_privacy}
	Consider a network of $N$ agents represented by a sequence of time-varying directed graphs $\{\mathcal{G}(k)=(\mathcal{V}, \, \mathcal{E}(k))\}$ which satisfy Assumptions \ref{assumption_strongly_connected}, \ref{assumption_interval_bound}, \ref{assumption_out_degree}, \ref{assumption_initial_states}, and \ref{assumption_collude}. Under the proposed Algorithm 1, if at some time instant $0\leq k^* \leq K$, agent $i\notin \mathcal{A}$ has an in-neighbor or out-neighbor $l$ not belonging to $\mathcal{A}$, then under $\mathcal{I}_s^* \triangleq \{\Delta s_{mn}(k) \, | \, (m,\,n) \in \mathcal{E}(k), (m,\,n)\neq (i,\,l), (m,\,n)\neq (l,\,i), k=0,1,\ldots,K\}$, the joint probability distributions of $s_i(K+1)$ and $s_l(K+1)$ are identical for any two different initial states $\mathbf{x}^0, \tilde{\mathbf{x}}^0 \in [a, \, b]^N$ subject to $x_i^0+x_l^0=\tilde{x}_i^0 + \tilde{x}_l^0$ and $x_j^0=\tilde{x}_j^0$ for $j\in \mathcal{V}\setminus \{i,l\}$.
\end{Lemma}

{\it Proof}: According to (\ref{Algorithm_I_s_update}), we can rewrite the update rule of $s_i(k)$ as follows
\begin{equation}\label{eqn_s_update_1} 
	\begin{aligned}
		& s_i(k+1) \\
		= & {\rm frac}\Big(\sum_{j\in \mathcal{N}_i^{in}(k)} \Delta s_{ij}(k) + s_i(k) - \sum_{j\in \mathcal{N}_i^{out}(k)} \Delta s_{ji}(k) \Big)
	\end{aligned}
\end{equation}
for $k \leq K$. We stack all variables $\Delta s_{mn}(k)$ into a vector $\boldsymbol{\Delta}\mathbf{s}(k)$ according to indices of edges in $\mathcal{E}(k)$, namely, the index of $\Delta s_{mn}(k)$ is determined by the index of the edge $(m, \, n)$ in $\mathcal{E}(k)$. Then we can further rewrite (\ref{eqn_s_update_1}) in the following more compact form:
\begin{equation}\label{eqn_s_update_2}
	\begin{aligned}
		\mathbf{s}(k+1)={\rm frac}\big(\mathbf{s}(k)+\mathbf{C}(k) \boldsymbol{\Delta}\mathbf{s}(k)\big)
	\end{aligned} 
\end{equation}
for $k \leq K$, where $\mathbf{C}(k)$ is the incidence matrix for graph $\mathcal{G}(k)$ at iteration $k$.

Define the subsets of agents $\mathcal{H}$ and $\mathcal{R}$ as $\mathcal{H}=\{i,l\}$ and $\mathcal{R}=\mathcal{V}\setminus (\mathcal{H}\cup\mathcal{A})$, respectively. Let $N_{\mathcal{A}}=|\mathcal{A}|$ and $N_{\mathcal{R}}=|\mathcal{R}|$ represent the number of agents in $\mathcal{A}$ and $\mathcal{R}$, respectively. It is clear that $\mathcal{H}$, $\mathcal{A}$, and $\mathcal{R}$ are disjoint, and $\mathcal{H} \cup \mathcal{A} \cup \mathcal{R} = \mathcal{V}$ holds. For graph $\mathcal{G}(k)=(\mathcal{V},\mathcal{E}(k))$, we define the subgraph $\mathcal{G_H}(k)$ as $\mathcal{G_H}(k)=(\mathcal{H},\mathcal{E_H}(k))$ where $\mathcal{E_H}(k) \subset \mathcal{E}(k)$ is the set of edges entirely within $\mathcal{H}$. The union of subgraphs $\mathcal{G_H}(k)$ for iterations $k=0,1,\ldots, K$ is further denoted as $\mathcal{G}_{\mathcal{H}}^*= \bigcup\limits_{k=0}^K \mathcal{G_H}(k) = (\mathcal{H}, \mathcal{E}_{\mathcal{H}}^*)$, where $\mathcal{E}_{\mathcal{H}}^*=\bigcup\limits_{k=0}^K \mathcal{E_H}(k)$ is the union of edge sets $\mathcal{E_H}(k)$ for iterations $k=0,1,\ldots, K$. Denote the edge set $\mathcal{E_A}(k)$ as the collection of edges associated with all agents in $\mathcal{A}$ at iteration $k$, i.e., $\mathcal{E_A}(k)=\{(m,\,n) \, | \, (m,\,n) \in \mathcal{E}(k), m \ \textnormal{or} \ n \in \mathcal{A}\}$. Then the set of remaining edges not belonging to $\mathcal{E_H}(k)$ or $\mathcal{E_A}(k)$ can be denoted as $\mathcal{E_R}(k)=\mathcal{E}(k)\setminus (\mathcal{E_H}(k) \cup \mathcal{E_A}(k))$, which is a collection of edges that are either entirely within $\mathcal{R}$ or between $\mathcal{R}$ and $\mathcal{H}$. Let $E_{\mathcal{H}}(k)=|\mathcal{E_H}(k)|$, $E_{\mathcal{A}}(k)=|\mathcal{E_A}(k)|$, and $E_{\mathcal{R}}(k)=|\mathcal{E_R}(k)|$ represent the number of edges in $\mathcal{E_H}(k)$, $\mathcal{E_A}(k)$, and $\mathcal{E_R}(k)$, respectively. Note that $\mathcal{E_H}(k)$, $\mathcal{E_A}(k)$, and $\mathcal{E_R}(k)$ are disjoint, and we always have $\mathcal{E_H}(k) \cup \mathcal{E_A}(k) \cup \mathcal{E_R}(k) = \mathcal{E}(k)$. Without loss of generality, we can partition the state vector $\mathbf{s}(k)$ as $\mathbf{s}(k)=[\mathbf{s}_{\mathcal{H}}(k)^T \ \mathbf{s}_{\mathcal{A}}(k)^T \ \mathbf{s}_{\mathcal{R}}(k)^T]^T$ where $\mathbf{s}_{\mathcal{H}}(k)$, $\mathbf{s}_{\mathcal{A}}(k)$, and $\mathbf{s}_{\mathcal{R}}(k)$ denote the state vectors of agents in $\mathcal{H}$, $\mathcal{A}$, and $\mathcal{R}$, respectively. Then we can further rewrite the update rule of $\mathbf{s}(k)$ for $k \leq K$ as
\begin{equation}\label{eqn_s_update_3}
	\begin{aligned}
		& \begin{bmatrix}
			\mathbf{s}_{\mathcal{H}}(k+1)\\
			\mathbf{s}_{\mathcal{A}}(k+1)\\
			\mathbf{s}_{\mathcal{R}}(k+1)
		\end{bmatrix}
		={\rm frac}\Bigg(
		\begin{bmatrix}
			\mathbf{s}_{\mathcal{H}}(k)\\
			\mathbf{s}_{\mathcal{A}}(k)\\
			\mathbf{s}_{\mathcal{R}}(k)
		\end{bmatrix} \\
		& \quad \ +
		\begin{bmatrix}
			\mathbf{C}_{\mathcal{H}}(k) & \mathbf{C}_{\mathcal{A}}^1(k) & \mathbf{C}_{\mathcal{R}}^1(k) \\
			\mathbf{0}_{N_{\mathcal{A}} \times E_{\mathcal{H}}(k)} & \mathbf{C}_{\mathcal{A}}^2(k) & \mathbf{0}_{N_{\mathcal{A}} \times E_{\mathcal{R}}(k)} \\
			\mathbf{0}_{N_{\mathcal{R}} \times E_{\mathcal{H}}(k)} & \mathbf{C}_{\mathcal{A}}^3(k) & \mathbf{C}_{\mathcal{R}}^2(k) \\
		\end{bmatrix}
		\begin{bmatrix}
			\boldsymbol{\Delta}\mathbf{s}_{\mathcal{H}}(k)\\
			\boldsymbol{\Delta}\mathbf{s}_{\mathcal{A}}(k)\\
			\boldsymbol{\Delta}\mathbf{s}_{\mathcal{R}}(k)
		\end{bmatrix}
		\Bigg)
	\end{aligned}
\end{equation}
where $\mathbf{C}_{\mathcal{H}}(k)$ is the incidence matrix for subgraph $\mathcal{G_H}(k)=(\mathcal{H},\mathcal{E_H}(k))$, and $\boldsymbol{\Delta}\mathbf{s}_{\mathcal{H}}(k)$, $\boldsymbol{\Delta}\mathbf{s}_{\mathcal{A}}(k)$, and $\boldsymbol{\Delta}\mathbf{s}_{\mathcal{R}}(k)$ are vectors stacking $\Delta s_{mn}(k)$ corresponding to edge sets $\mathcal{E_H}(k)$, $\mathcal{E_A}(k)$, and $\mathcal{E_R}(k)$, respectively. It is obvious that $\boldsymbol{\Delta}\mathbf{s}_{\mathcal{A}}(k)$ is completely known to agents in $\mathcal{A}$ since every element in $\boldsymbol{\Delta}\mathbf{s}_{\mathcal{A}}(k)$ is either sent or received by the agents in $\mathcal{A}$. From (\ref{eqn_s_update_3}), we can further obtain
\begin{equation}\label{eqn_s_update_4}
	\begin{aligned}
		& \begin{bmatrix}
			\mathbf{s}_{\mathcal{H}}(K+1)\\
			\mathbf{s}_{\mathcal{A}}(K+1)\\
			\mathbf{s}_{\mathcal{R}}(K+1)
		\end{bmatrix}
		={\rm frac}\Bigg(
		\begin{bmatrix}
			\mathbf{s}_{\mathcal{H}}(0)\\
			\mathbf{s}_{\mathcal{A}}(0)\\
			\mathbf{s}_{\mathcal{R}}(0)
		\end{bmatrix}\\
		& \qquad \qquad +
		\begin{bmatrix}
			\mathbf{C}_{\mathcal{H}}^* & \mathbf{C}_{\mathcal{A}}^{1*} & \mathbf{C}_{\mathcal{R}}^{1*}\\
			\mathbf{0}_{N_{\mathcal{A}} \times E_{\mathcal{H}}^*} & \mathbf{C}_{\mathcal{A}}^{2*} & \mathbf{0}_{N_{\mathcal{A}} \times E_{\mathcal{R}}^*} \\
			\mathbf{0}_{N_{\mathcal{R}} \times E_{\mathcal{H}}^*} & \mathbf{C}_{\mathcal{A}}^{3*} & \mathbf{C}_{\mathcal{R}}^{2*}
		\end{bmatrix}
		\begin{bmatrix}
			\boldsymbol{\Delta}\mathbf{s}_{\mathcal{H}}^*\\
			\boldsymbol{\Delta}\mathbf{s}_{\mathcal{A}}^*\\
			\boldsymbol{\Delta}\mathbf{s}_{\mathcal{R}}^*
		\end{bmatrix}
		\Bigg)
	\end{aligned} 
\end{equation}
where $E_{\mathcal{H}}^*=\sum_{k=0}^{K}E_{\mathcal{H}}(k)$, $E_{\mathcal{R}}^*=\sum_{k=0}^{K}E_{\mathcal{R}}(k)$, and
\begin{equation}
	\begin{aligned}
		\mathbf{C}_{\mathcal{H}}^* & =
		\begin{bmatrix}
			\mathbf{C}_{\mathcal{H}}(0) & \cdots & \mathbf{C}_{\mathcal{H}}(K)\\
		\end{bmatrix} \\
		\mathbf{C}_{\mathcal{A}}^{t*} & =
		\begin{bmatrix}
			\mathbf{C}_{\mathcal{A}}^t(0) & \cdots & \mathbf{C}_{\mathcal{A}}^t(K)\\
		\end{bmatrix}, \ t=1,2,3 \\
		\mathbf{C}_{\mathcal{R}}^{t*} & =
		\begin{bmatrix}
			\mathbf{C}_{\mathcal{R}}^t(0) & \cdots & \mathbf{C}_{\mathcal{R}}^t(K)\\
		\end{bmatrix}, \ t=1,2 \\
		\boldsymbol{\Delta}\mathbf{s}_{\mathcal{H}}^* & =
		\begin{bmatrix}
			\boldsymbol{\Delta}\mathbf{s}_{\mathcal{H}}(0)^T & \cdots & \boldsymbol{\Delta}\mathbf{s}_{\mathcal{H}}(K)^T\\
		\end{bmatrix}^T \\
		\boldsymbol{\Delta}\mathbf{s}_{\mathcal{A}}^* & =
		\begin{bmatrix}
			\boldsymbol{\Delta}\mathbf{s}_{\mathcal{A}}(0)^T & \cdots & \boldsymbol{\Delta}\mathbf{s}_{\mathcal{A}}(K)^T\\
		\end{bmatrix}^T \\
		\boldsymbol{\Delta}\mathbf{s}_{\mathcal{R}}^* & =
		\begin{bmatrix}
			\boldsymbol{\Delta}\mathbf{s}_{\mathcal{R}}(0)^T & \cdots & \boldsymbol{\Delta}\mathbf{s}_{\mathcal{R}}(K)^T\\
		\end{bmatrix}^T \\
	\end{aligned} 
\end{equation}
It is worth noting that $\mathbf{C}_{\mathcal{H}}^*$ is the incidence matrix for graph $\mathcal{G}_{\mathcal{H}}^* = \bigcup\limits_{k=0}^K \mathcal{G_H}(k)$. From (\ref{eqn_s_update_4}), we have
\begin{equation}\label{eqn_s_update_5}
	\begin{aligned}
		& \mathbf{s}_{\mathcal{H}}(K+1) \\
		= \ & {\rm frac}\big(\mathbf{s}_{\mathcal{H}}(0) + \mathbf{C}_{\mathcal{H}}^* \boldsymbol{\Delta}\mathbf{s}_{\mathcal{H}}^* + \mathbf{C}_{\mathcal{A}}^{1*}
		\boldsymbol{\Delta}\mathbf{s}_{\mathcal{A}}^* + \mathbf{C}_{\mathcal{R}}^{1*}
		\boldsymbol{\Delta}\mathbf{s}_{\mathcal{R}}^*\big)
	\end{aligned} 
\end{equation}
Denoting $\mathbf{v}=[v_1 \, v_2]^T$ as
\begin{equation}
	\begin{aligned}
		\mathbf{v}\triangleq{\rm frac}\big(\mathbf{C}_{\mathcal{H}}^* \boldsymbol{\Delta}\mathbf{s}_{\mathcal{H}}^*\big)
	\end{aligned} 
\end{equation}
next we prove that $\mathbf{v}$ is uniformly distributed in $[0, \, 1) \times [0, \, 1)$ subject to ${\rm frac}(v_1+v_2)=0$ if at some time instant $0\leq k^* \leq K$, agent $i$ has an in-neighbor or out-neighbor $l$ not belonging to $\mathcal{A}$.

Given $\mathcal{H}=\{i,l\}$, if at some time instant $0\leq k^* \leq K$, agent $i$ has an in-neighbor or out-neighbor $l$ not belonging to $\mathcal{A}$, the columns of $\mathbf{C}_{\mathcal{H}}^*$ are either $[1 \ -1]^T$ ($l$ is an in-neighbor of agent $i$) or $[-1 \ 1]^T$ ($l$ is an out-neighbor of agent $i$). Denote the $j$-th column of $\mathbf{C}_{\mathcal{H}}^*$ as $\mathbf{c}_j^*$, then $\mathbf{c}_j^*$ can be expressed as 
\begin{equation}
	\begin{aligned}
		\mathbf{c}_j^*= r_{j} \mathbf{c}_1^*
	\end{aligned} 
\end{equation}
where $r_{j}\in \{1, -1\}$ is the corresponding coefficient. Defining $\mathbf{r}$ as $\mathbf{r}=[r_1 \cdots r_{E_{\mathcal{H}}^*}]$, one can obtain 
\begin{equation}
	\begin{aligned}
		\mathbf{v} & ={\rm frac}\big(\mathbf{C}_{\mathcal{H}}^* \boldsymbol{\Delta}\mathbf{s}_{\mathcal{H}}^*\big) = {\rm frac}\big(\mathbf{c}_1^* \mathbf{r} \boldsymbol{\Delta}\mathbf{s}_{\mathcal{H}}^*\big) \\
		& = {\rm frac}\big(\mathbf{c}_1^* \cdot {\rm frac}(\mathbf{r} \boldsymbol{\Delta}\mathbf{s}_{\mathcal{H}}^*)\big)
	\end{aligned} 
\end{equation}
Since all the elements in $\boldsymbol{\Delta}\mathbf{s}_{\mathcal{H}}^*$ are independent of each other and uniformly distributed in $[0, \, 1)$, we have that ${\rm frac}(\mathbf{r} \boldsymbol{\Delta}\mathbf{s}_{\mathcal{H}}^*)$ is uniformly distributed in $[0, \, 1)$. As $\mathbf{c}_1^*$ is either $[1 \ -1]^T$ or $[-1 \ 1]^T$, we have that $\mathbf{v}$ is uniformly distributed in $[0, \, 1) \times [0, \, 1)$ subject to ${\rm frac}(v_1+v_2)=0$.

From (\ref{eqn_s_update_5}), we have $\mathbf{s}_{\mathcal{H}}(K+1)={\rm frac}\big(\mathbf{s}_{\mathcal{H}}(0) + \mathbf{v} + \mathbf{C}_{\mathcal{A}}^{1*} \boldsymbol{\Delta}\mathbf{s}_{\mathcal{A}}^* + \mathbf{C}_{\mathcal{R}}^{1*} \boldsymbol{\Delta}\mathbf{s}_{\mathcal{R}}^*\big)$. Further taking into account the fact that $\mathbf{v}$ is uniformly distributed in $[0, \, 1) \times [0, \, 1)$ subject to ${\rm frac}(v_1 + v_2) =0$ if at some time instant $0\leq k^* \leq K$, agent $i$ has an in-neighbor or out-neighbor $l$ not belonging to $\mathcal{A}$, we can obtain that $\mathbf{s}_{\mathcal{H}}(K+1)$ is also uniformly distributed in $[0, \, 1) \times [0, \, 1)$ subject to ${\rm frac}(\mathbf{1}^T\mathbf{s}_{\mathcal{H}}(K+1)) = {\rm frac}(\mathbf{1}^T\mathbf{s}_{\mathcal{H}}(0) + \mathbf{1}^T \mathbf{C}_{\mathcal{A}}^{1*} \boldsymbol{\Delta}\mathbf{s}_{\mathcal{A}}^* + \mathbf{1}^T \mathbf{C}_{\mathcal{R}}^{1*} \boldsymbol{\Delta}\mathbf{s}_{\mathcal{R}}^*)$.

Now we analyze the probability distributions of $\mathbf{s}_{\mathcal{H}}(K+1)$ under different initial conditions of agent $i$. For any two different initial conditions $\mathbf{x}^0, \, \tilde{\mathbf{x}}^0 \in [a, \, b]^N$ subject to $x_i^0+x_l^0=\tilde{x}_i^0+\tilde{x}_l^0$ and $x_j^0=\tilde{x}_j^0$ for $j\in \mathcal{V}\setminus \{i,l\}$, one can obtain $s_i(0)+s_l(0) = \tilde{s}_i(0)+\tilde{s}_l(0)$. Note that all the elements in $\boldsymbol{\Delta}\mathbf{s}_{\mathcal{A}}^*$ and $\boldsymbol{\Delta}\mathbf{s}_{\mathcal{R}}^*$ belong to the set $\mathcal{I}_s^* = \{\Delta s_{mn}(k) \, | \, (m,\,n) \in \mathcal{E}(k), (m,\,n)\neq (i,\,l), (m,\,n)\neq (l,\,i), k=0,1,\ldots,K\}$. Thus, given $\mathcal{I}_s^*$, both $\mathbf{s}_{\mathcal{H}}(K+1)$ and $\tilde{\mathbf{s}}_{\mathcal{H}}(K+1)$ are uniformly distributed in $[0, \, 1) \times [0, \, 1)$ subject to ${\rm frac}(\mathbf{1}^T\mathbf{s}_{\mathcal{H}}(K+1)) = {\rm frac}(\mathbf{1}^T\mathbf{s}_{\mathcal{H}}(0) + \mathbf{1}^T \mathbf{C}_{\mathcal{A}}^{1*} \boldsymbol{\Delta}\mathbf{s}_{\mathcal{A}}^* + \mathbf{1}^T \mathbf{C}_{\mathcal{R}}^{1*} \boldsymbol{\Delta}\mathbf{s}_{\mathcal{R}}^*) = {\rm frac}(\mathbf{1}^T\tilde{\mathbf{s}}_{\mathcal{H}}(0) + \mathbf{1}^T \mathbf{C}_{\mathcal{A}}^{1*} \boldsymbol{\Delta}\mathbf{s}_{\mathcal{A}}^* + \mathbf{1}^T \mathbf{C}_{\mathcal{R}}^{1*} \boldsymbol{\Delta}\mathbf{s}_{\mathcal{R}}^*) = {\rm frac}(\mathbf{1}^T \tilde{\mathbf{s}}_{\mathcal{H}}(K+1))$. Therefore, given $\mathcal{I}_s^*$, the probability distributions of $\mathbf{s}_{\mathcal{H}}(K+1)$ under different $\mathbf{x}^0$ and $\tilde{\mathbf{x}}^0$ are identical when $x_j^0=\tilde{x}_j^0$ for $j\in \mathcal{V}\setminus \{i,l\}$ and $x_i^0+x_l^0=\tilde{x}_i^0+\tilde{x}_l^0$ hold for some agent $l$ that is an in-neighbor or out-neighbor of agent $i$ at some time instant $0\leq k^* \leq K$ but does not belong to $\mathcal{A}$. \hfill{$\blacksquare$}

Now we are in position to present our main confidentiality result.

\begin{Theorem}\label{theorem_preserve_privacy}
	Consider a network of $N$ agents represented by a sequence of time-varying directed graphs $\{\mathcal{G}(k)=(\mathcal{V}, \, \mathcal{E}(k))\}$ which satisfy Assumptions \ref{assumption_strongly_connected}, \ref{assumption_interval_bound}, \ref{assumption_out_degree}, \ref{assumption_initial_states}, and \ref{assumption_collude}. Under the proposed Algorithm 1, the confidentiality of agent $i\notin \mathcal{A}$ can be preserved against $\mathcal{A}$ if at some time instant $0\leq k^* \leq K$, agent $i$ has an in-neighbor or out-neighbor $l$ not belonging to $\mathcal{A}$.
\end{Theorem}

{\it Proof}: To show that the confidentiality of agent $i$ can be protected, we have to show that under different initial values of agent $i$, the probability distributions are identical for honest-but-curious agents $\mathcal{A}$' information set. It suffices to prove that the probability distributions of information sets accessible to $\mathcal{A}$ are identical for all $\mathbf{x}^0, \, \tilde{\mathbf{x}}^0 \in [a, \, b]^N$ subject to $x_i^0+x_l^0=\tilde{x}_i^0+\tilde{x}_l^0$ and $x_j^0=\tilde{x}_j^0$ for $j\in \mathcal{V}\setminus \{i,l\}$ where agent $l$ is an in-neighbor or out-neighbor of agent $i$ at some time instant $0\leq k^* \leq K$ and does not belong to $\mathcal{A}$. Note that under such $\mathbf{x}^0$ and $\tilde{\mathbf{x}}^0$, the sum of the initial states of all nodes not in $\mathcal{A}$ keeps unchanged, i.e., $\sum_{ j\in \mathcal{V}\setminus\mathcal{A}} x_j^0 = \sum_{ j\in \mathcal{V}\setminus\mathcal{A}} \tilde{x}_j^0$. Given $s_i(0)= \frac{1}{N^2} + \frac{(N-2)(x_i^0-a)}{(b-a)N^2}$, this is equivalent to proving that the probability distributions of information sets accessible to $\mathcal{A}$ are identical for all $\mathbf{s}(0), \, \tilde{\mathbf{s}}(0) \in [\frac{1}{N^2}, \, \frac{N-1}{N^2}]^N$ subject to $s_i(0) + s_l(0) = \tilde{s}_i(0) + \tilde{s}_l(0)$ and $s_j(0)=\tilde{s}_j(0)$ for $j\in \mathcal{V}\setminus \{i,l\}$. Denoting $\mathcal{I}_s(k)$ and $\mathcal{I}_w(k)$ as
\begin{equation}
	\begin{aligned}
		\mathcal{I}_s(k) & = \{ \mathbf{s}_{\mathcal{A}}(k), \boldsymbol{\Delta}\mathbf{s}_{\mathcal{A}}(k)\} \\
		\mathcal{I}_w(k) & = \{ \mathbf{w}_{\mathcal{A}}(k), \boldsymbol{\Delta}\mathbf{w}_{\mathcal{A}}(k)\}
	\end{aligned} 
\end{equation}
where $\mathbf{s}_{\mathcal{A}}(k)$ and $\mathbf{w}_{\mathcal{A}}(k)$ are state vectors of agents in $\mathcal{A}$, and $\boldsymbol{\Delta}\mathbf{s}_{\mathcal{A}}(k)$ and $\boldsymbol{\Delta}\mathbf{w}_{\mathcal{A}}(k)$ are augmented vectors of $\Delta s_{mn}(k)$ and $\Delta w_{mn}(k)$ corresponding to edge set $\mathcal{E_A}(k)$, respectively. We can see that agents in $\mathcal{A}$ have access to both $\mathcal{I}_s(k)$ and $\mathcal{I}_w(k)$ at each iteration $k$. Further denote $\mathcal{I}_1^s$, $\mathcal{I}_1^w$, $\mathcal{I}_2^s$, $\mathcal{I}_2^w$, $\mathcal{I}_1$, and $\mathcal{I}_2$ as
\begin{equation}
	\begin{aligned}
		\mathcal{I}_1^s & = \bigcup\limits_{k=0}^K \mathcal{I}_s(k) \qquad \quad \, \mathcal{I}_1^w = \bigcup\limits_{k=0}^K \mathcal{I}_w(k) \cup \{ \mathbf{w}(0)\} \\
		\mathcal{I}_2^s & = \bigcup\limits_{k=K+1}^{\infty} \mathcal{I}_s(k) \qquad \mathcal{I}_2^w = \bigcup\limits_{k=K+1}^{\infty} \mathcal{I}_w(k) \\
		\mathcal{I}_1 & = \mathcal{I}_1^s \cup \mathcal{I}_1^w \qquad \quad \ \ \mathcal{I}_2 = \mathcal{I}_2^s \cup \mathcal{I}_2^w 
	\end{aligned} 
\end{equation}
According to Algorithm 1, $\mathcal{I}\triangleq\mathcal{I}_1 \cup \mathcal{I}_2$ represents all the information accessible to $\mathcal{A}$. We denote the conditional probability density of $\mathcal{I}$ given $\mathbf{s}(0)$ as $f(\mathcal{I} \, | \, \mathbf{s}(0))$. Thus, to show that the confidentiality of agent $i$ can be protected, it is equivalent to proving $f(\mathcal{I} \, | \, \mathbf{s}(0)) = f(\mathcal{I} \, | \, \tilde{\mathbf{s}}(0))$ for all $\mathbf{s}(0), \, \tilde{\mathbf{s}}(0) \in [\frac{1}{N^2}, \, \frac{N-1}{N^2}]^N$ subject to $s_i(0) + s_l(0) = \tilde{s}_i(0) + \tilde{s}_l(0)$ and $s_j(0)=\tilde{s}_j(0)$ for $j\in \mathcal{V}\setminus \{i,l\}$.

Since 
\begin{equation}
	\begin{aligned}
		f(\mathcal{I} \, | \, \mathbf{s}(0)) = f(\mathcal{I}_1, \mathcal{I}_2 \, | \, \mathbf{s}(0)) = f(\mathcal{I}_1 \, | \, \mathbf{s}(0)) f(\mathcal{I}_2 \, | \, \mathcal{I}_1, \mathbf{s}(0))
	\end{aligned} 
\end{equation}
holds, to show $f(\mathcal{I} \, | \, \mathbf{s}(0)) = f(\mathcal{I} \, | \, \tilde{\mathbf{s}}(0))$, it suffices to prove that
\begin{equation}
	\begin{aligned}
		f(\mathcal{I}_1 \, | \, \mathbf{s}(0)) = f(\mathcal{I}_1 \, | \, \tilde{\mathbf{s}}(0))
	\end{aligned} 
\end{equation}
and
\begin{equation}
	\begin{aligned}
		f(\mathcal{I}_2 \, | \, \mathcal{I}_1, \mathbf{s}(0))= f(\mathcal{I}_2 \, | \, \mathcal{I}_1, \tilde{\mathbf{s}}(0))
	\end{aligned} 
\end{equation}
hold for any $\mathbf{s}(0), \, \tilde{\mathbf{s}}(0) \in [\frac{1}{N^2}, \, \frac{N-1}{N^2}]^N$ subject to $s_i(0) + s_l(0) = \tilde{s}_i(0) + \tilde{s}_l(0)$ and $s_j(0)=\tilde{s}_j(0)$ for $j\in \mathcal{V}\setminus \{i,l\}$.

We first show $f(\mathcal{I}_1 \, | \, \mathbf{s}(0)) = f(\mathcal{I}_1 \, | \, \tilde{\mathbf{s}}(0)$. Since $\mathcal{I}_1^w$ is independent of $\mathcal{I}_1^s$ and $\mathbf{s}(0)$, one can obtain $f(\mathcal{I}_1 \, | \, \mathbf{s}(0)) = f(\mathcal{I}_1^s, \mathcal{I}_1^w \, | \, \mathbf{s}(0))= f(\mathcal{I}_1^w) f(\mathcal{I}_1^s \, | \, \mathbf{s}(0))$. From (\ref{eqn_s_update_3}), we can see that given $\mathbf{s}_{\mathcal{A}}(0)$, $\mathcal{I}_1^s$ is independent of $\mathbf{s}_{\mathcal{H}}(0)$ and $\mathbf{s}_{\mathcal{R}}(0)$, where $\mathcal{H}=\{i,l\}$ and $\mathcal{R}=\mathcal{V}\setminus (\mathcal{H} \cup \mathcal{A})$. So $f(\mathcal{I}_1^s \, | \, \mathbf{s}(0))= f(\mathcal{I}_1^s \, | \, \mathbf{s}_{\mathcal{A}}(0), \mathbf{s}_{\mathcal{H}}(0), \mathbf{s}_{\mathcal{R}}(0)) = f(\mathcal{I}_1^s \, | \, \mathbf{s}_{\mathcal{A}}(0))$ holds. Therefore, we have
\begin{equation}
	\begin{aligned}
		f(\mathcal{I}_1 \, | \, \mathbf{s}(0)) & = f(\mathcal{I}_1^w) f(\mathcal{I}_1^s \, | \, \mathbf{s}_{\mathcal{A}}(0)) = f(\mathcal{I}_1^w) f(\mathcal{I}_1^s \, | \, \tilde{\mathbf{s}}_{\mathcal{A}}(0)) \\
		& = f(\mathcal{I}_1 \, | \, \tilde{\mathbf{s}}(0))
	\end{aligned} 
\end{equation}
where we used $\mathbf{s}_{\mathcal{A}}(0)=\tilde{\mathbf{s}}_{\mathcal{A}}(0)$ in the derivation.

To show $f(\mathcal{I}_2 \, | \, \mathcal{I}_1, \mathbf{s}(0))= f(\mathcal{I}_2 \, | \, \mathcal{I}_1, \tilde{\mathbf{s}}(0))$, it suffices to prove
\begin{equation}
	\begin{aligned}
		& f(\mathcal{I}_2, \mathbf{s}(K+1), \mathbf{w}(K+1) \, | \, \mathcal{I}_1, \mathbf{s}(0))\\
		= \ & f(\mathcal{I}_2, \mathbf{s}(K+1), \mathbf{w}(K+1) \, | \, \mathcal{I}_1, \tilde{\mathbf{s}}(0))
	\end{aligned} 
\end{equation}
Given $\mathbf{s}(K+1)$ and $\mathbf{w}(K+1)$, $\mathcal{I}_2$ is independent of $\mathcal{I}_1$, which further leads to
\begin{equation}
	\begin{aligned}
		& f(\mathcal{I}_2, \mathbf{s}(K+1), \mathbf{w}(K+1) \, | \, \mathcal{I}_1, \mathbf{s}(0)) \\
		= & f(\mathcal{I}_2 \, | \, \mathbf{s}(K+1), \mathbf{w}(K+1), \mathcal{I}_1, \mathbf{s}(0)) \\ 
		& \times f(\mathbf{s}(K+1), \mathbf{w}(K+1) \, | \, \mathcal{I}_1, \mathbf{s}(0)) \\
		= & f(\mathcal{I}_2 \, | \, \mathbf{s}(K+1), \mathbf{w}(K+1)) \\
		& \times f(\mathbf{s}(K+1), \mathbf{w}(K+1) \, | \, \mathcal{I}_1, \mathbf{s}(0)) \\
		= & f(\mathcal{I}_2 \, | \, \mathbf{s}(K+1), \mathbf{w}(K+1)) \\
		& \times f(\mathbf{s}(K+1) \, | \, \mathbf{w}(K+1), \mathcal{I}_1, \mathbf{s}(0)) f(\mathbf{w}(K+1) \, | \, \mathcal{I}_1, \mathbf{s}(0))
	\end{aligned} 
\end{equation}
Further taking into account the facts that 1) $\mathbf{s}(K+1)$ is conditionally independent of $\mathbf{w}(K+1)$ and $\mathcal{I}_1^w$ given $\mathcal{I}_1^s$ and $\mathbf{s}(0)$; and 2) $\mathbf{w}(K+1)$ is conditionally independent of $\mathcal{I}_1^s$ and $\mathbf{s}(0)$ given $\mathcal{I}_1^w$, one can obtain
\begin{equation}
	\begin{aligned}
		& f(\mathcal{I}_2, \mathbf{s}(K+1), \mathbf{w}(K+1) \, | \, \mathcal{I}_1, \mathbf{s}(0)) \\
		= & f(\mathcal{I}_2 \, | \, \mathbf{s}(K+1), \mathbf{w}(K+1)) f(\mathbf{s}(K+1) \, | \, \mathcal{I}_1^s, \mathbf{s}(0)) \\
		& \times f(\mathbf{w}(K+1) \, | \, \mathcal{I}_1^w)
	\end{aligned} 
\end{equation}
Similarly, one can also obtain
\begin{equation}
	\begin{aligned}
		& f(\mathcal{I}_2, \mathbf{s}(K+1), \mathbf{w}(K+1) \, | \, \mathcal{I}_1, \tilde{\mathbf{s}}(0)) \\
		= & f(\mathcal{I}_2 \, | \, \mathbf{s}(K+1), \mathbf{w}(K+1)) f(\mathbf{s}(K+1) \, | \, \mathcal{I}_1^s, \tilde{\mathbf{s}}(0)) \\
		& \times f(\mathbf{w}(K+1) \, | \, \mathcal{I}_1^w)
	\end{aligned} 
\end{equation}
Therefore, to show $f(\mathcal{I}_2 \, | \, \mathcal{I}_1, \mathbf{s}(0))= f(\mathcal{I}_2 \, | \, \mathcal{I}_1, \tilde{\mathbf{s}}(0))$, it suffices to prove $f(\mathbf{s}(K+1) \, | \, \mathcal{I}_1^s, \mathbf{s}(0))=f(\mathbf{s}(K+1) \, | \, \mathcal{I}_1^s, \tilde{\mathbf{s}}(0))$. Denote $\mathcal{I}_{\mathcal{R}}^s$ as $\mathcal{I}_{\mathcal{R}}^s \triangleq \{\boldsymbol{\Delta} \mathbf{s}_{\mathcal{R}}(k) \, | \, k=0,1,\ldots,K \}$. To prove $f(\mathbf{s}(K+1) \, | \, \mathcal{I}_1^s, \mathbf{s}(0))=f(\mathbf{s}(K+1) \, | \, \mathcal{I}_1^s, \tilde{\mathbf{s}}(0))$, we only need to prove 
\begin{equation}
	\begin{aligned}
		f(\mathbf{s}(K+1), \mathcal{I}_{\mathcal{R}}^s \, | \, \mathcal{I}_1^s, \mathbf{s}(0))=f(\mathbf{s}(K+1), \mathcal{I}_{\mathcal{R}}^s \, | \, \mathcal{I}_1^s, \tilde{\mathbf{s}}(0))
	\end{aligned} 
\end{equation}
From (\ref{eqn_s_update_4}), we can obtain the following facts: 1) $\mathbf{s}_{\mathcal{H}}(K+1)$ is conditionally independent of $\mathbf{s}_{\mathcal{A}}(K+1)$ and $\mathbf{s}_{\mathcal{R}}(K+1)$ given $\mathcal{I}_{\mathcal{R}}^s$, $\mathcal{I}_1^s$, and $\mathbf{s}(0)$; and 2) $\mathbf{s}_{\mathcal{A}}(K+1)$ and $\mathbf{s}_{\mathcal{R}}(K+1)$ are conditionally independent of $\mathbf{s}_{\mathcal{H}}(0)$ given $\mathcal{I}_{\mathcal{R}}^s$, $\mathcal{I}_1^s$, $\mathbf{s}_{\mathcal{A}}(0)$, and $\mathbf{s}_{\mathcal{R}}(0)$. Taking into account these facts, we have
\begin{equation}
	\begin{aligned}
		& f(\mathbf{s}(K+1), \mathcal{I}_{\mathcal{R}}^s \, | \, \mathcal{I}_1^s, \mathbf{s}(0)) \\
		= & f(\mathbf{s}_{\mathcal{H}}(K+1), \mathbf{s}_{\mathcal{A}}(K+1), \mathbf{s}_{\mathcal{R}}(K+1), \mathcal{I}_{\mathcal{R}}^s \, | \, \mathcal{I}_1^s, \mathbf{s}(0)) \\
		= & f(\mathbf{s}_{\mathcal{H}}(K+1) \, | \, \mathbf{s}_{\mathcal{A}}(K+1), \mathbf{s}_{\mathcal{R}}(K+1), \mathcal{I}_{\mathcal{R}}^s, \mathcal{I}_1^s, \mathbf{s}(0)) \\
		& \times f(\mathbf{s}_{\mathcal{A}}(K+1), \mathbf{s}_{\mathcal{R}}(K+1) \, | \, \mathcal{I}_{\mathcal{R}}^s, \mathcal{I}_1^s, \mathbf{s}(0)) 
		 f(\mathcal{I}_{\mathcal{R}}^s \, | \, \mathcal{I}_1^s,\mathbf{s}(0)) \\
		= & f(\mathbf{s}_{\mathcal{H}}(K+1) \, | \, \mathcal{I}_{\mathcal{R}}^s, \mathcal{I}_1^s, \mathbf{s}(0)) \\
		& \times f(\mathbf{s}_{\mathcal{A}}(K+1), \mathbf{s}_{\mathcal{R}}(K+1) \, | \, \mathcal{I}_{\mathcal{R}}^s, \mathcal{I}_1^s, \mathbf{s}_{\mathcal{A}}(0), \mathbf{s}_{\mathcal{R}}(0)) f(\mathcal{I}_{\mathcal{R}}^s) \\ 
	\end{aligned} 
\end{equation}
where in the derivation we used the independence between $\mathcal{I}_{\mathcal{R}}^s$ and $\{\mathcal{I}_1^s, \mathbf{s}(0)\}$. Similarly, one can obtain
\begin{equation}
	\begin{aligned}
		& f(\mathbf{s}(K+1), \mathcal{I}_{\mathcal{R}}^s \, | \, \mathcal{I}_1^s,\tilde{\mathbf{s}}(0)) \\
		= & f(\mathbf{s}_{\mathcal{H}}(K+1) \, | \, \mathcal{I}_{\mathcal{R}}^s, \mathcal{I}_1^s, \tilde{\mathbf{s}}(0))\\
		& \times f(\mathbf{s}_{\mathcal{A}}(K+1), \mathbf{s}_{\mathcal{R}}(K+1) \, | \, \mathcal{I}_{\mathcal{R}}^s, \mathcal{I}_1^s, \tilde{\mathbf{s}}_{\mathcal{A}}(0), \tilde{\mathbf{s}}_{\mathcal{R}}(0)) f(\mathcal{I}_{\mathcal{R}}^s) \\ 
	\end{aligned} 
\end{equation}
From Lemma \ref{lemma_privacy}, we have that if at some time instant $0\leq k^* \leq K$, agent $i$ has an in-neighbor or out-neighbor $l$ not belonging to $\mathcal{A}$, $f(\mathbf{s}_{\mathcal{H}}(K+1) \, | \, \mathcal{I}_s^*, \mathbf{s}(0)) = f(\mathbf{s}_{\mathcal{H}}(K+1) \, | \, \mathcal{I}_s^*, \tilde{\mathbf{s}}(0))$ holds, where $\mathcal{I}_s^* = \{\Delta s_{mn}(k) \, | \, (m,\,n) \in \mathcal{E}(k), (m,\,n)\neq (i,\,l), (m,\,n)\neq (l,\,i), k=0,1,\ldots,K\} = \{\Delta s_{mn}(k) \, | \, (m,\,n) \in \mathcal{E_A}(k) \cup \mathcal{E_R}(k), k=0,1,\ldots,K\}$ is the collection of all elements $\Delta s_{mn}(k)$ in $\boldsymbol{\Delta}\mathbf{s}_{\mathcal{A}}(k)$ and $\boldsymbol{\Delta}\mathbf{s}_{\mathcal{R}}(k)$ from iteration $0$ to iteration $K$. Given that $\boldsymbol{\Delta}\mathbf{s}_{\mathcal{A}}(k) \in \mathcal{I}_1^s$ and $\boldsymbol{\Delta}\mathbf{s}_{\mathcal{R}}(k) \in \mathcal{I}_{\mathcal{R}}^s$ hold for $k=0,1,\ldots,K$, we further have $f(\mathbf{s}_{\mathcal{H}}(K+1) \, | \, \mathcal{I}_{\mathcal{R}}^s, \mathcal{I}_1^s, \mathbf{s}(0)) = f(\mathbf{s}_{\mathcal{H}}(K+1) \, | \, \mathcal{I}_{\mathcal{R}}^s, \mathcal{I}_1^s, \tilde{\mathbf{s}}(0)$. Based on $\mathbf{s}_{\mathcal{A}}(0)=\tilde{\mathbf{s}}_{\mathcal{A}}(0)$ and $\mathbf{s}_{\mathcal{R}}(0)=\tilde{\mathbf{s}}_{\mathcal{R}}(0)$, we have $f(\mathbf{s}(K+1), \mathcal{I}_{\mathcal{R}}^s \, | \, \mathcal{I}_1^s, \mathbf{s}(0)) = f(\mathbf{s}(K+1), \mathcal{I}_{\mathcal{R}}^s \, | \, \mathcal{I}_1^s, \tilde{\mathbf{s}}(0))$, implying $f(\mathcal{I}_2 \, | \, \mathcal{I}_1, \mathbf{s}(0)) =f(\mathcal{I}_2 \, | \, \mathcal{I}_1, \tilde{\mathbf{s}}(0))$ if at some time instant $0\leq k^* \leq K$, agent $i$ has an in-neighbor or out-neighbor $l$ not belonging to $\mathcal{A}$.

Therefore, we have $f(\mathcal{I} \, | \, \mathbf{s}(0)) =f(\mathcal{I} \, | \, \tilde{\mathbf{s}}(0))$ for any $\mathbf{s}(0), \, \tilde{\mathbf{s}}(0) \in [\frac{1}{N^2}, \, \frac{N-1}{N^2}]^N$ subject to $s_i(0) + s_l(0) = \tilde{s}_i(0) + \tilde{s}_l(0)$ and $s_j(0)=\tilde{s}_j(0)$ for $j\in \mathcal{V}\setminus \{i,l\}$, meaning that the confidentiality of agent $i$ can be preserved if at some time instant $0\leq k^* \leq K$, agent $i$ has an in-neighbor or out-neighbor $l$ that does not belong to $\mathcal{A}$. \hfill{$\blacksquare$}

\begin{Remark}\label{remark_difference_our_CNS}
	Compared with our previous work \cite{Huan2018CNS} which defines privacy as the positivity of probability that adversaries' accessible information set being the same under two different initial states, in this work we significantly improved our confidential results by proving that the probability distributions of information sets are identical under different initial states, meaning that the initial states are perfectly indistinguishable from the viewpoint of adversaries.
\end{Remark}

\begin{Remark}\label{remark_time_space_difference}
	It is worth noting that although the confidential approach in \cite{pilet2019robust} looks similar to ours (both protocols employ random values in the first several iterations), they are in fact significantly different. More specifically, to guarantee the accuracy of average consensus, not only does the confidential protocol in \cite{pilet2019robust} require pairwise local averaging of exchanged random values in the first several iterations, but it also needs to compensate the errors induced by the random values immediately after the first several iterations, which is equivalent to introducing time-correlated additive random noises to agents' states. To the contrary, our confidential approach exchanges time-uncorrelated random values for iterations $k \leq K$. To ensure consensus accuracy, each agent $i$ uses a carefully-designed $\Delta s_{ii}(k)$ to compensate the errors induced by the random values at each iteration $k \leq K$. Therefore, the random values used in our approach are space-correlated. As shown in Theorem \ref{theorem_preserve_privacy}, the space-correlated randomness can make the probability distributions of information sets accessible to adversaries identical under different initial states and hence achieves information-theoretic privacy, which is stronger than the confidentiality achieved using time-correlated noises in \cite{pilet2019robust}.
\end{Remark}

Next we proceed to show that if the conditions in Theorem \ref{theorem_preserve_privacy} are not met, then the confidentiality of agent $i$ may be breached. 

\begin{Theorem}\label{theorem_no_privacy}
Consider a network of $N$ agents represented by a sequence of time-varying directed graphs $\{\mathcal{G}(k)=(\mathcal{V}, \, \mathcal{E}(k))\}$ which satisfy Assumptions \ref{assumption_strongly_connected}, \ref{assumption_interval_bound}, \ref{assumption_out_degree}, \ref{assumption_initial_states}, and \ref{assumption_collude}. Under the proposed Algorithm 1, the confidentiality of agent $i\notin \mathcal{A}$ cannot be preserved against $\mathcal{A}$ if all in-neighbors and out-neighbors of agent $i$ belong to $\mathcal{A}$, i.e., $\{\mathcal{N}_i^{in}(k)\cup \mathcal{N}_i^{out}(k) \, \big| \, k \geq 0 \} \subset \mathcal{A}$. In fact, when $\{\mathcal{N}_i^{in}(k)\cup \mathcal{N}_i^{out}(k) \, \big| \, k \geq 0 \} \subset \mathcal{A}$ is true, the initial value $x_i^0$ of agent $i$ can be uniquely determined by honest-but-curious agents in $\mathcal{A}$.
\end{Theorem}

{\it Proof}: From (\ref{eqn_s_update_1}) we have
\begin{equation}\label{theorem_3_1}
	\begin{aligned}
		& {\rm frac}\big(s_i(k+1)-s_i(k) \big) \\
		= & {\rm frac}\Big(\sum_{n\in \mathcal{N}_i^{in}(k)} \Delta s_{in}(k) - \sum_{m\in \mathcal{N}_i^{out}(k)} \Delta s_{mi}(k) \Big)
	\end{aligned}
\end{equation}
for $k \leq K$ and further
\begin{equation}\label{theorem_3_2}
	\begin{aligned}
		& {\rm frac}\big(s_i(K+1)-s_i(0) \big) \\
		= & {\rm frac}\bigg(\sum_{k=0}^{K} \Big[\sum_{n\in \mathcal{N}_i^{in}(k)} \Delta s_{in}(k) - \sum_{m\in \mathcal{N}_i^{out}(k)} \Delta s_{mi}(k) \Big] \bigg)
	\end{aligned}
\end{equation}
Since $s_i(0) \in [\frac{1}{N^2}, \, \frac{N-1}{N^2}]$ holds, one can obtain
\begin{equation}\label{theorem_3_3}
	\begin{aligned}
		s_i(0) = & {\rm frac}\bigg(s_i(K+1) \\
		& - \sum_{k=0}^{K} \Big[\sum_{n\in \mathcal{N}_i^{in}(k)} \Delta s_{in}(k) - \sum_{m\in \mathcal{N}_i^{out}(k)} \Delta s_{mi}(k) \Big] \bigg)
	\end{aligned}
\end{equation}

According to the requirements in Algorithm 1, we have $\Delta s_{ii}(k) + \sum_{m\in \mathcal{N}_i^{out}(k)}{\Delta s_{mi}(k)}=s_i(k)$ for $k \geq K+1$ and $\Delta w_{ii}(k) + \sum_{m\in \mathcal{N}_i^{out}(k)}{\Delta w_{mi}(k)}=w_i(k)$ for $k \geq 0$. Plugging these relationships into (\ref{Algorithm_I_s_update}) and (\ref{Algorithm_I_w_update}), we can obtain
\begin{equation}\label{eqn_theorem_3_4}
	\begin{aligned}
		& s_i(k+1)- s_i(k) = \\
		& \qquad \sum\limits_{n\in \mathcal{N}_i^{in}(k)}{\Delta s_{in}(k)} - \sum\limits_{m\in \mathcal{N}_i^{out}(k)}{\Delta s_{mi}(k)}\\
	\end{aligned}
\end{equation}
for $k \geq K+1$ and
\begin{equation}\label{eqn_theorem_3_5}
	\begin{aligned}
		& w_i(k+1)- w_i(k)= \\
		& \qquad \sum\limits_{n\in \mathcal{N}_i^{in}(k)}{\Delta w_{in}(k)} - \sum\limits_{m\in \mathcal{N}_i^{out}(k)}{\Delta w_{mi}(k)} \\
	\end{aligned}
\end{equation}
for $k \geq 0$, which further lead to
\begin{equation}\label{eqn_theorem_3_6}
	\begin{aligned}
		& s_i(k)- s_i(K+1)= \\
		& \qquad \sum\limits_{l=K+1}^{k-1} \Big[ \sum\limits_{n\in \mathcal{N}_i^{in}(k)}{\Delta s_{in}(l)} - \sum\limits_{m\in \mathcal{N}_i^{out}(k)}{\Delta s_{mi}(l)} \Big]
	\end{aligned}
\end{equation}
for $k \geq K+1$ and
\begin{equation}\label{eqn_theorem_3_7}
	\begin{aligned}
		& w_i(k)- w_i(0)= \\
		& \qquad \sum\limits_{l=0}^{k-1} \Big[ \sum\limits_{n\in \mathcal{N}_i^{in}(k)}{\Delta w_{in}(l)} - \sum\limits_{m\in \mathcal{N}_i^{out}(k)}{\Delta w_{mi}(l)} \Big]\\
	\end{aligned}
\end{equation}
for $k \geq 0$.

Under Assumption \ref{assumption_collude}, if $\{\mathcal{N}_i^{out}(k)\cup \mathcal{N}_i^{in}(k) \, \big| \, k \geq 0 \} \subset \mathcal{A}$ is true, then all terms on the right-hand side of (\ref{eqn_theorem_3_6}) and (\ref{eqn_theorem_3_7}) are known to the honest-but-curious agents in $\mathcal{A}$. Further taking into account $w_i(0)=1$, we have that agents in $\mathcal{A}$ can uniquely determine $w_i(k)$ for all $k$. Under Assumption \ref{assumption_strongly_connected} and $\{\mathcal{N}_i^{out}(k)\cup \mathcal{N}_i^{in}(k) \, \big| \, k \geq 0 \} \subset \mathcal{A}$, there must exist at least one agent $j \in \mathcal{A}$ such that $(j, \, i) \in \mathcal{E}_{\infty}$ is true. This is because otherwise graph $(\mathcal{V}, \, \mathcal{E}_{\infty})$ is not strongly connected, which does not satisfy Assumption \ref{assumption_strongly_connected}. According to Assumption \ref{assumption_interval_bound}, agent $i$ directly communicates with agent $j\in \mathcal{A}$ at least once in every $T$ consecutive time instants. So there must exist $k' \geq K+1$ at which agent $i$ directly communicates with agent $j$, i.e., agent $i$ sends $\Delta s_{ji}(k')$ and $\Delta w_{ji}(k')$ to agent $j$ at iteration $k'$. As $j \in \mathcal{A}$, every honest-but-curious agent in $\mathcal{A}$ has access to $\Delta s_{ji}(k')$ and $\Delta w_{ji}(k')$. So agents in $\mathcal{A}$ can easily infer $s_i(k')$ by using the following relationship
\begin{equation}\label{agent_i_s_5_A}
	s_i(k') = \frac{\Delta s_{ji}(k')}{\Delta w_{ji}(k')} w_i(k') = \frac{p_{ji}(k')s_i(k')}{p_{ji}(k')w_i(k')} w_i(k')
\end{equation}
and then determine the value of $s_i(K+1)$ using (\ref{eqn_theorem_3_6}). Further making use of (\ref{theorem_3_3}), agents in $\mathcal{A}$ can infer the value of $s_i(0)$, and then uniquely determine the initial value of agent $i$ using $x_i^0=\frac{b-a}{N-2}(N^2s_i(0)-1)+a$. Therefore, the confidentiality of agent $i\notin \mathcal{A}$ cannot be preserved against $\mathcal{A}$ if all in-neighbors and out-neighbors of agent $i$ belong to $\mathcal{A}$. \hfill{$\blacksquare$}

\begin{Remark}\label{remark_topology_requirements}
	It is worth noting that in confidential average consensus, topology requirements such as the ones in Theorem \ref{theorem_preserve_privacy} are widely used. In fact, to guarantee both accuracy and confidentiality, \cite{manitara2013privacy, he2018privacy, mo2017privacy, charalambous2019privacy, gupta2019statistical, pilet2019robust, altafini2019dynamical, pequito2014design, ridgley2019simple, fang2018secure, ruan2019secure, wang2019privacy} all rely on similar topology requirements.
\end{Remark}

\begin{Remark}\label{remark_different_from_other_papers}
	Our algorithm can protect the confidentiality of an agent even when all its neighbors interact (at least one does not collude) with adversaries, which is not allowed in \cite{mo2017privacy} and \cite{manitara2013privacy}.
\end{Remark}

\begin{Remark}\label{remark_tradeoff_privacy_convergence}
	From the above analysis, we know that introducing randomness into interaction dynamics by each agent $i$ for $k\leq K$ is key to protect confidentiality against honest-but-curious agents. It is worth noting that compared with the conventional push-sum approach which does not take confidentiality into consideration, the introduced randomness in our approach has no influence on the convergence rate $\gamma$. However, the randomness does delay the convergence process and hence leads to a trade-off between confidentiality preservation and convergence time. This is confirmed in our numerical simulations in Fig. \ref{convergence_error_figure}, which shows that convergence only initiates after $k = K+1$.
\end{Remark}

\begin{Remark}\label{remark_tradeoff_intermediate_privacy}	
	If an adversary can obtain side information, then a larger $K$ protects the confidentiality of more intermediate states $s_i(k)$ for $1\leq k\leq K$. This is because for $k \geq K+1$, $s_i(k)$ can be easily obtained by its out-neighbor $j$ due to the relationship $s_i(k)=w_i(k) \Delta s_{ji}(k)/\Delta w_{ji}(k)$ if side information about $w_i(k)$ is available to the adversary $j$. Therefore, although a larger $K$ leads to more delay in the convergence process, as discussed in Remark \ref{remark_tradeoff_privacy_convergence}, it can protect more intermediate states ($s_i(k)$ for $1 \leq k \leq K$) when an adversary can obtain side information. Of course, if side information is not of concern, a smaller $K$ is preferable to minimize the delay in the convergence process.
\end{Remark}

\begin{Remark}\label{remark_eavesdropper}
	Our algorithm can be extended to preserve confidentiality against external eavesdroppers wiretapping all communication links without compromising algorithmic accuracy by patching partially homomorphic encryption. More specifically, using public-key cryptosystems (e.g., Paillier \cite{paillier1999public}, RSA \cite{rivest1978method}, and ElGamal \cite{elgamal1985public}), each agent generates and floods its public key before the consensus iteration starts. Then in decentralized implementation, an agent encrypts its messages to be sent, which can be decrypted by a legitimate recipient without the help of any third party. Note that since public-key cryptosystems can only deal with integers, the final consensus result would be subject to a quantization error. However, as indicated in our previous work \cite{ruan2019secure}, the quantization error can be made arbitrarily small in implementation.
\end{Remark}

%%%%%%%%%%%%%%%%%%%%%%%%%%%%%%%%%%%%%%%%%%%%%%%
\section{Results Validation}

We conducted numerical simulations to verify the correctness and the effectiveness of our proposed approach.

\begin{figure}[h]
	\begin{center}
		\includegraphics[width=0.38\textwidth]{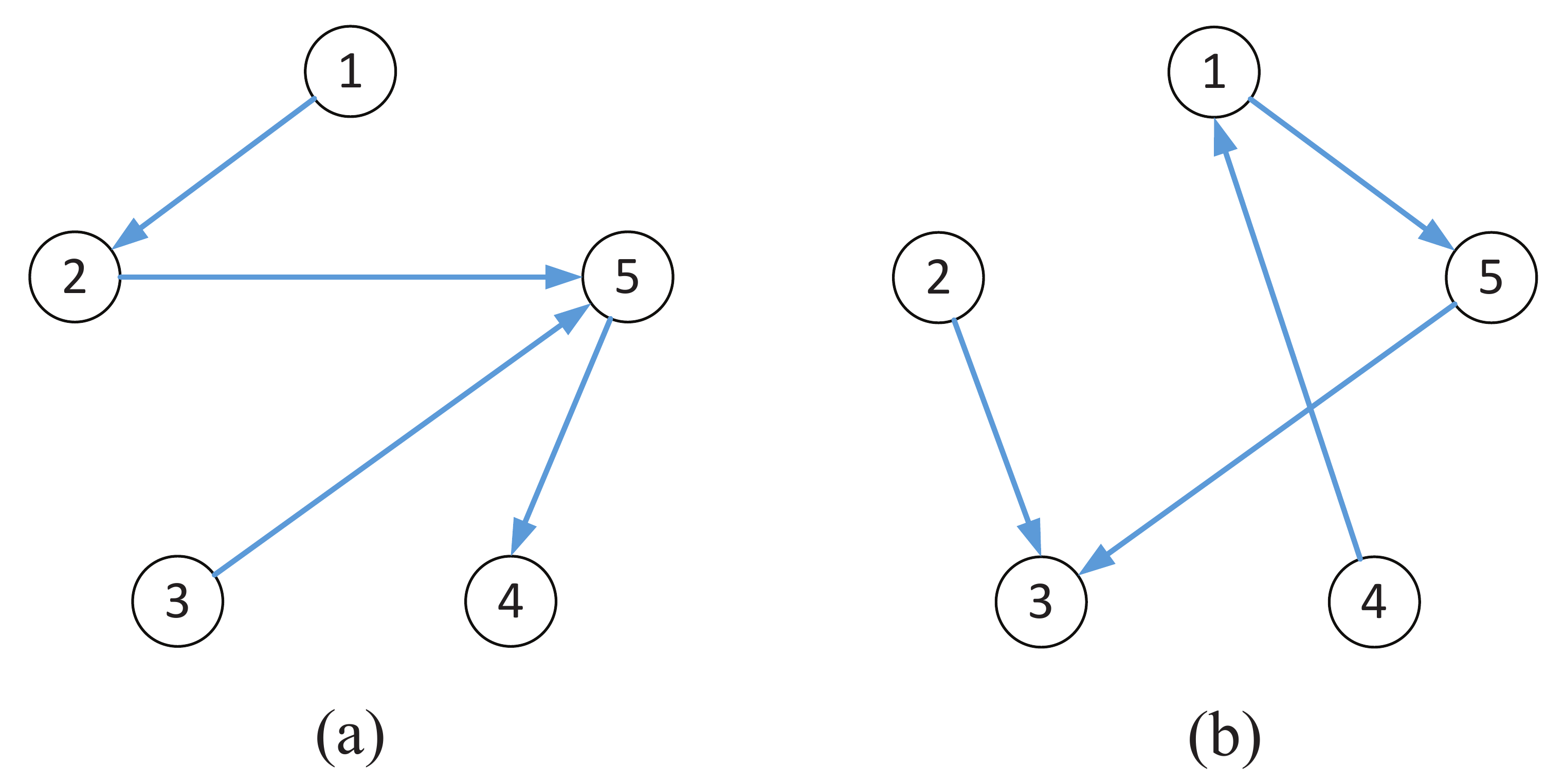}
	\end{center}
	\caption{A time-varying directed graph with $5$ agents.}
	\label{fig_graph_time_varying}
\end{figure}

\begin{figure}[h]
	\begin{center}
		\includegraphics[width=0.45\textwidth]{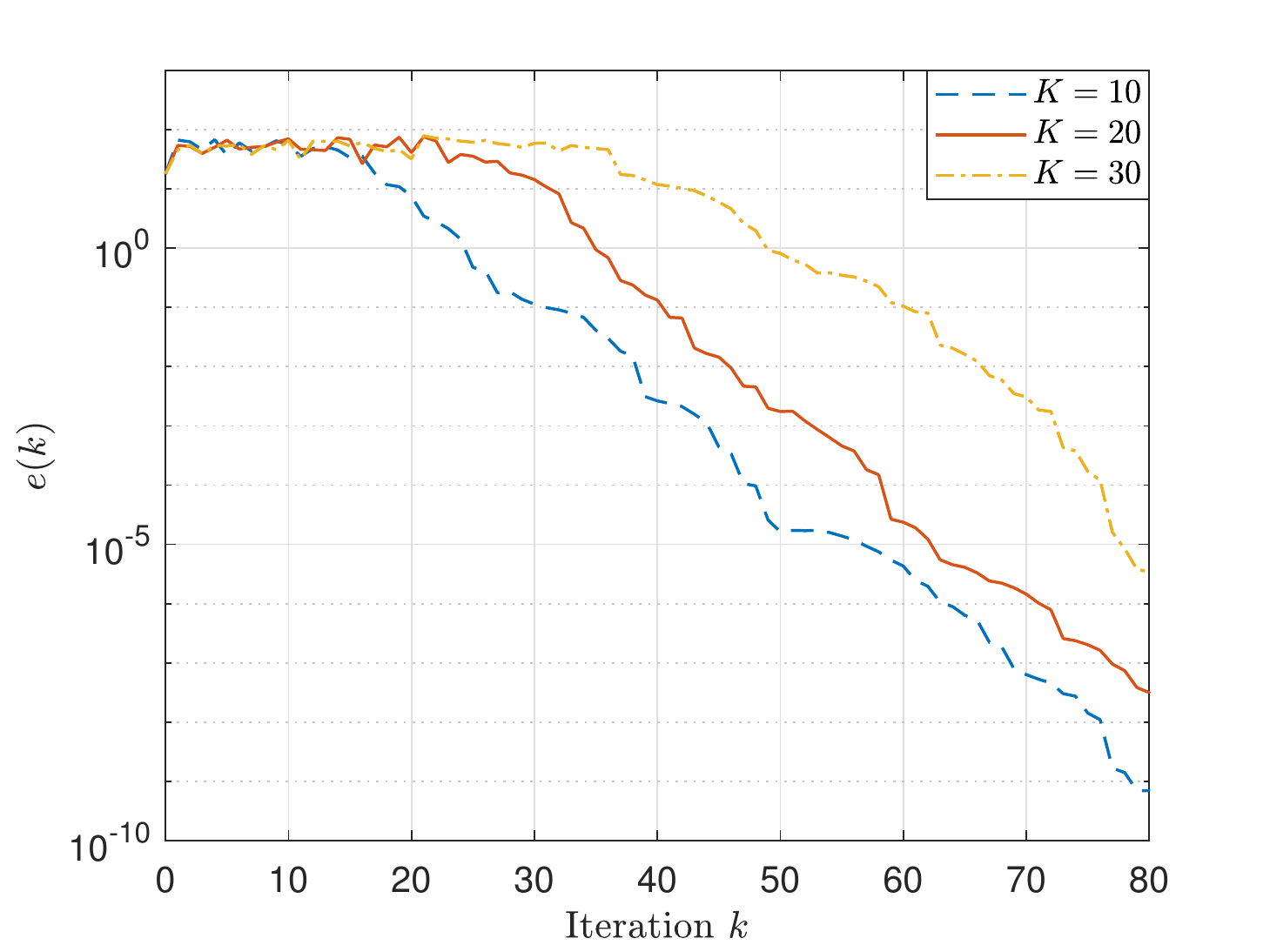}
	\end{center}
	\caption{The evolution of error $e(k)$ under different $K$.}
	\label{convergence_error_figure}
\end{figure}

We first evaluated our proposed Algorithm 1 under a network of $N=5$ agents interacting on a time-varying directed graph. More specifically, we used the interaction graph in Fig. \ref{fig_graph_time_varying}(a) when $k$ is even and Fig. \ref{fig_graph_time_varying}(b) when $k$ is odd. It can be verified that this time-varying directed graph satisfies Assumptions \ref{assumption_strongly_connected} and \ref{assumption_interval_bound}. Parameter $\varepsilon$ was set to $0.05$. The initial values $x_i^0$ for $i=1, \ldots, N$ were randomly chosen from $(-50, \, 50)$. We used $e(k)$ to measure the estimation error between the estimate $\pi_i(k)=s_i(k)/w_i(k)$ and the true average value $\bar{x}^0={\sum_{j=1}^{N}x_j^0}/{N}$ at iteration $k$, i.e.,
\begin{equation}\label{convergence_error}
\begin{aligned}
e(k)=\big\|\boldsymbol{\pi}(k) - \bar{x}^0 \mathbf{1} \big\|= \big( \sum\limits_{i=1}^{N}(\pi_i(k)-\bar{x}^0)^{2}\big)^{1/2}
\end{aligned}
\end{equation}
Three experiments were conducted with parameter $K$ being $10$, $20$, and $30$, respectively. The evolution of $e(k)$ is shown in Fig. \ref{convergence_error_figure}. It can be seen that $e(k)$ approached $0$, meaning that every agent converged to the average value $\bar{x}^0=\sum_{j=1}^{N}x_j^0/{N}$. From Fig. \ref{convergence_error_figure}, we can also see that Algorithm 1 did not start to converge in the first $K+1$ iterations due to the introduced randomness, which confirms our analysis in Remark \ref{remark_tradeoff_privacy_convergence}.

\begin{figure}[h]
	\begin{center}
		\includegraphics[width=0.45\textwidth]{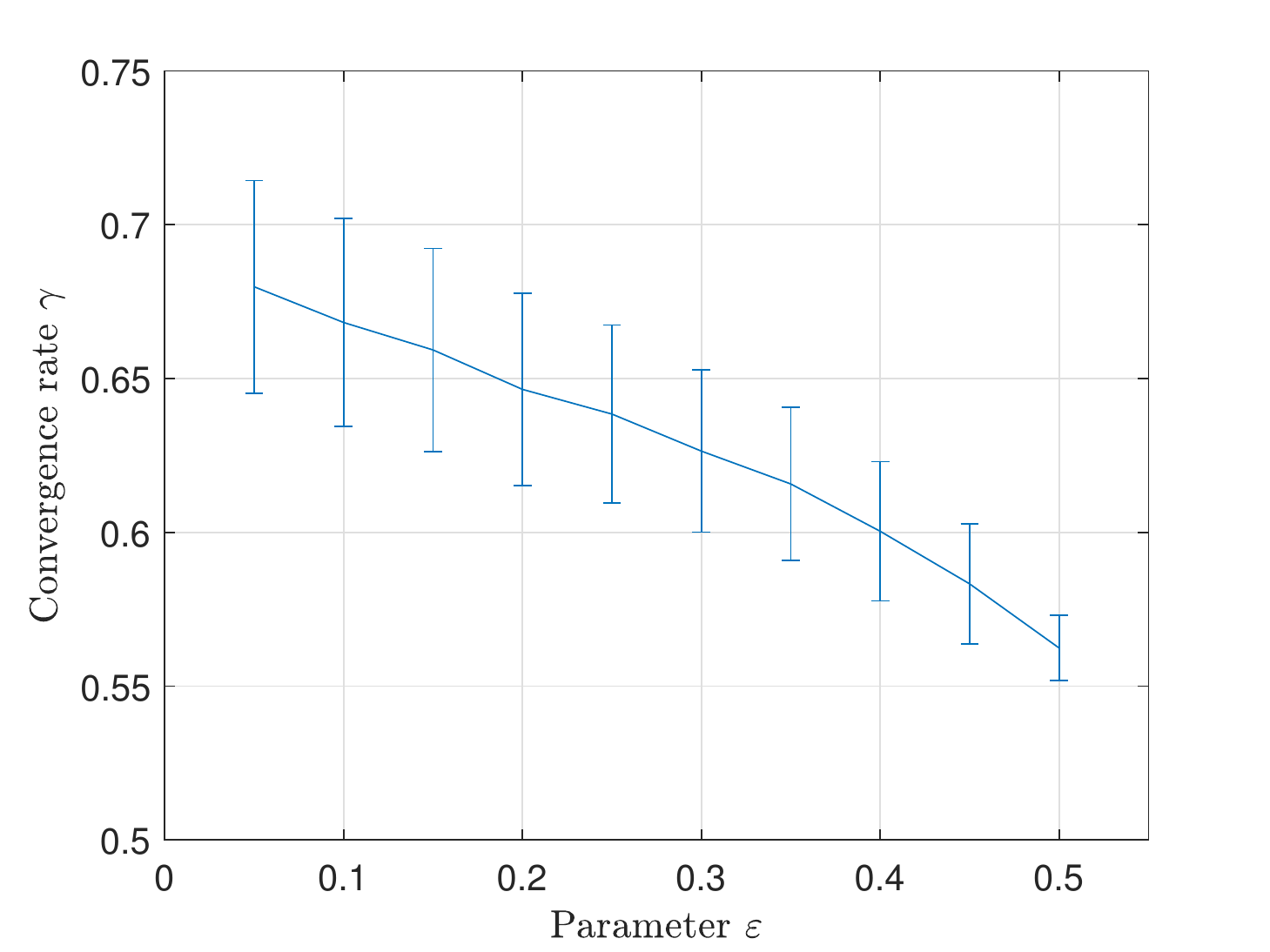}
	\end{center}
	\caption{The influence of $\varepsilon$ on the convergence rate $\gamma$.}
	\label{figure_epsilon_rho}
\end{figure}

We also evaluated the influence of parameter $\varepsilon$ on the convergence rate $\gamma$. The interaction graph was the same as above. $K$ was set to $10$. The simulation results are given in Fig. \ref{figure_epsilon_rho} where the mean and variance of $\gamma$ from $1,000$ runs of the algorithm are shown under different values of $\varepsilon$. Fig. \ref{figure_epsilon_rho} shows that as $\varepsilon$ increases, the convergence rate $\gamma$ decreases (i.e., the convergence speed increases), which confirms our analysis in Sec. III-B.

\begin{figure}[!h]
	\begin{center}
		\includegraphics[width=0.45\textwidth]{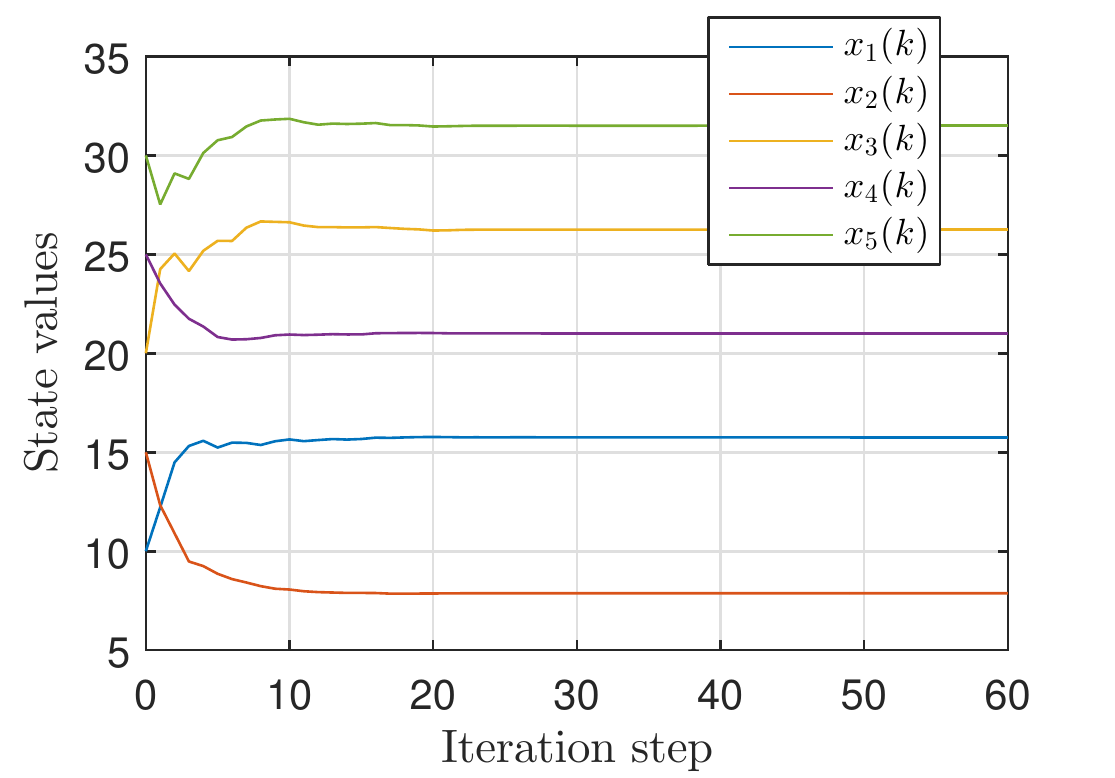}
	\end{center}
	\caption{The evolution of $x_i(k)$ under the approach in \cite{huang2012differentially}.}
	\label{comparison_Zhenqi_Huan_WPES}
\end{figure}

\begin{figure}[!h]
	\begin{center}
		\includegraphics[width=0.45\textwidth]{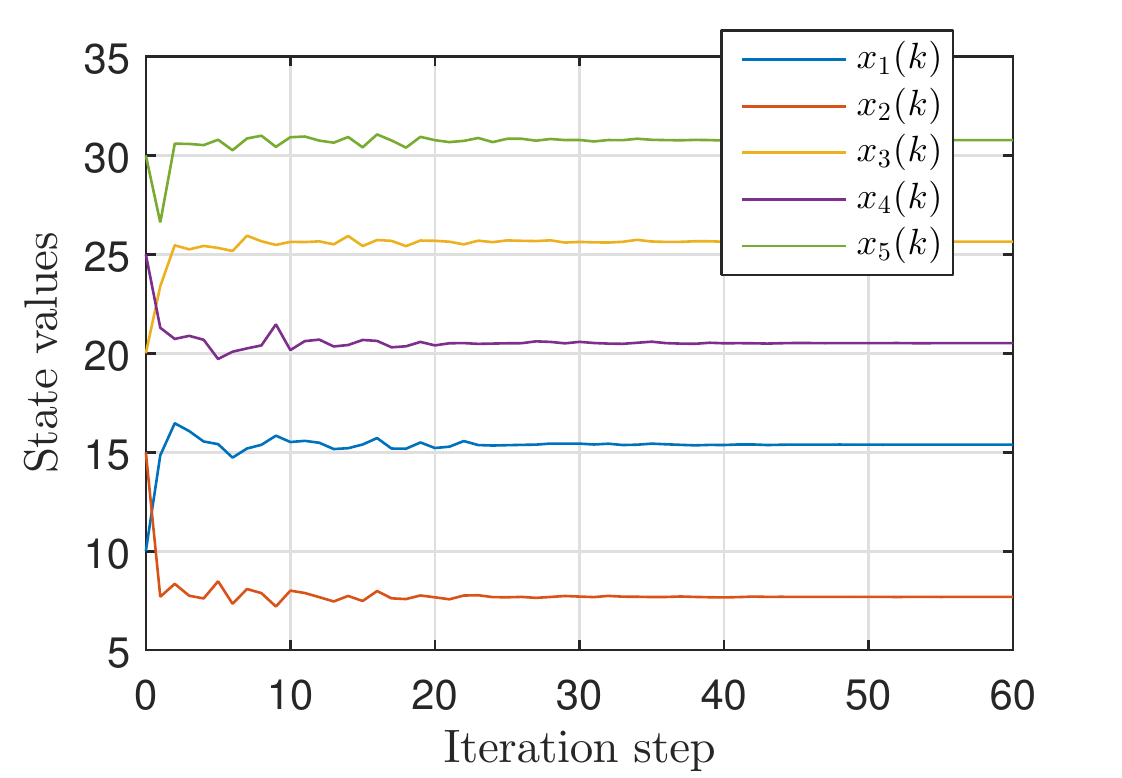}
	\end{center}
	\caption{The evolution of $x_i(k)$ under the approach in \cite{mo2017privacy}.}
	\label{comparison_Yilin_Mo}
\end{figure}

\begin{figure}[!h]
	\begin{center}
		\includegraphics[width=0.45\textwidth]{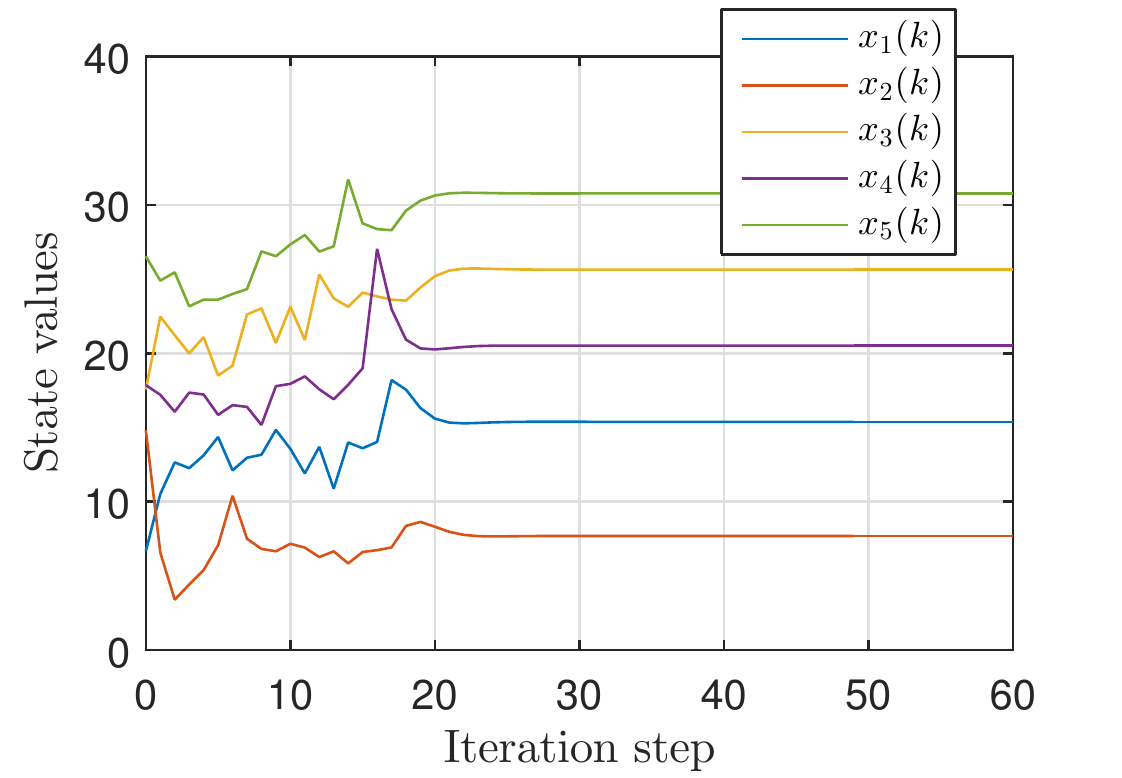}
	\end{center}
	\caption{The evolution of $x_i(k)$ under the approach in \cite{manitara2013privacy}.}
	\label{comparison_Nicolaos_Manitara_ECC}
\end{figure}

We then compared the proposed Algorithm 1 with existing data-obfuscation based approaches, more specifically, the differential-privacy based approach in \cite{huang2012differentially}, the decaying-noise approach in \cite{mo2017privacy}, and the finite-noise-sequence approach in \cite{manitara2013privacy}. Under the same setup as in the previous simulation, we chose the initial values as $\{10, 15, 20, 25, 30\}$, which led to an average value $20$. We adopted the weight matrix $\mathbf{W}$ from \cite{huang2012differentially}, i.e., the $ij$-th entry was $w_{ij}={1}/{(|\mathcal{N}_j^{out}|+1)}$ for $i\in \mathcal{N}_j^{out} \cup \{j\}$ and $w_{ij} = 0$ for $i\notin \mathcal{N}_j^{out}\cup \{j\}$. As the graph is directed and imbalanced, and does not meet the undirected or balanced assumption in \cite{huang2012differentially, manitara2013privacy, mo2017privacy}, all three approaches failed to achieve average consensus, as shown in the numerical simulation results in Fig. \ref{comparison_Zhenqi_Huan_WPES}, Fig. \ref{comparison_Yilin_Mo}, and Fig. \ref{comparison_Nicolaos_Manitara_ECC}, respectively.

\begin{figure}[!h]
	\begin{center}
		\includegraphics[width=0.45\textwidth]{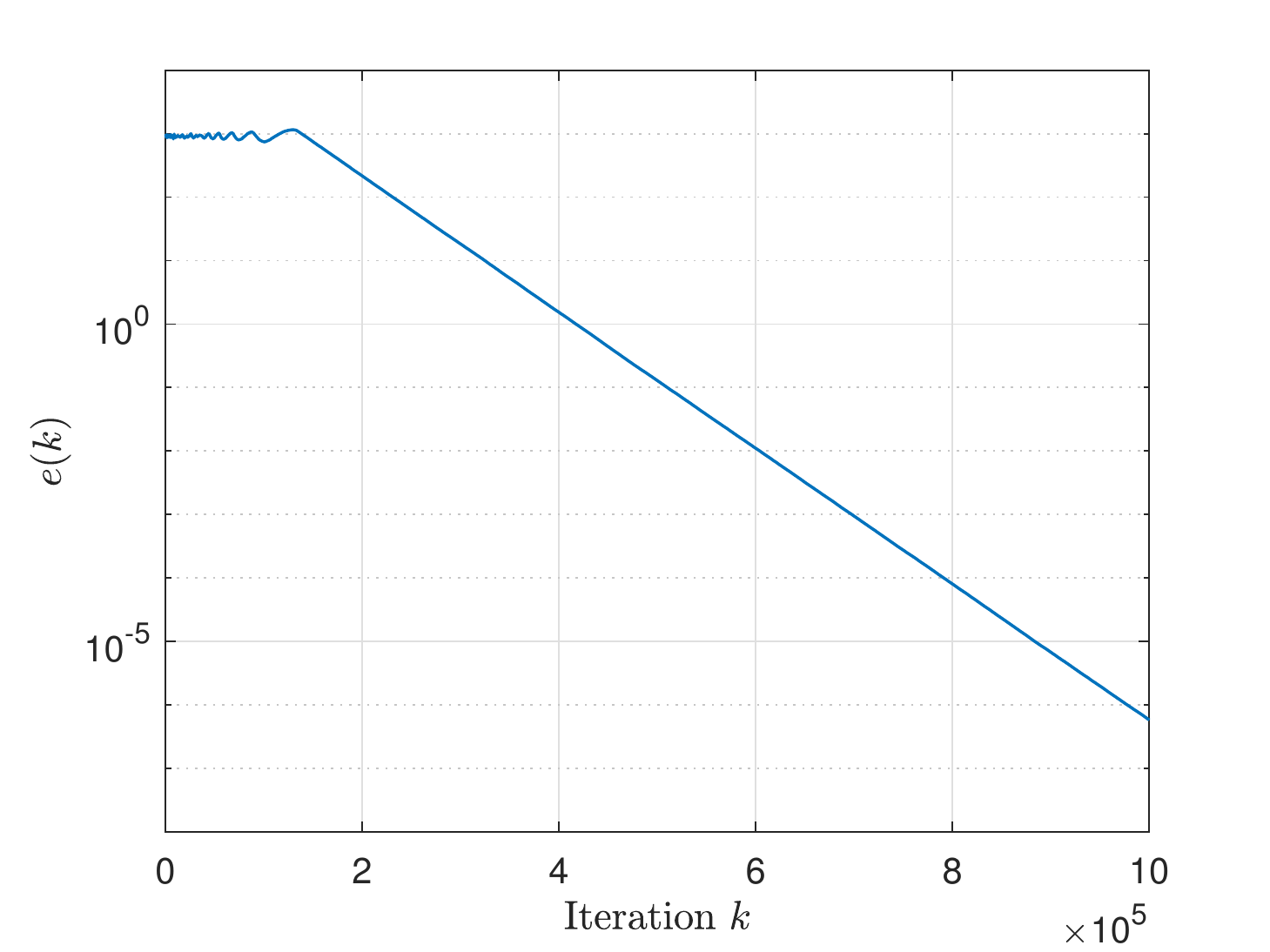}
	\end{center}
	\caption{The evolution of error $e(k)$ in a network of $N=1,000$ agents.}
	\label{fig_evolution_1000_agents}
\end{figure}

Finally, we conducted numerical simulations to verify the scalability of our proposed Algorithm 1 using a network of $N=1,000$ agents. At every iteration $k$, each agent $i$ was assumed to have three out-neighbors, i.e.,
\begin{equation}
	\mathcal{N}_i^{out}(k)=\left\{
	\begin{aligned}
		& \left\{\overline{i}+1,\overline{i+1}+1,\overline{i+2}+1\right\} \qquad \textnormal{if} \ k \ \textnormal{is even}\\
		& \left\{\overline{i-2}+1, \overline{i-3}+1,\overline{i-4}+1\right\} \ \textnormal{if} \ k \ \textnormal{is odd}\\
	\end{aligned}\right.
\end{equation}
where the superscript ``$\bar{\quad}$'' represents modulo operation on $N$, i.e., $\overline{i} \triangleq i \mod N$. The initial values $x_i^0$ for $i=1,2,\ldots,N$ were randomly chosen from $(-50,\, 50)$. $\varepsilon$ and $K$ were set to $0.05$ and $10$, respectively. The evolution of estimation error $e(k)=\|\boldsymbol{\pi}(k)-\bar{x}^0 \mathbf{1}\|$ is shown in Fig. \ref{fig_evolution_1000_agents}. It can be seen that $e(k)$ converged to $0$, implying that our proposed algorithm can guarantee the convergence of all agents to the actual average value even when the network size is large.

%%%%%%%%%%%%%%%%%%%%%%%%%%%%%%%%%%%%%%%%%%%%%%%%%%%%%%%%%%%%%%%%%%%%%%%%

%%%%%%%%%%%%%%%%%%%%%%%%%%%%%%%%%%%%%%%%%%%%%%%
\section{Conclusions}

We proposed a confidential average consensus algorithm for time-varying directed graphs. In distinct difference from existing differential-privacy based approaches which enable confidentiality through compromising the accuracy of obtained consensus value, we leveraged the inherent robustness of average consensus to embed randomness in interaction dynamics, which guarantees confidentiality of participating agents without sacrificing the accuracy of average consensus. Finally, we provided numerical simulation results to confirm the effectiveness and efficiency of our proposed approach.

\bibliographystyle{IEEEtran}
\bibliography{abbr_bibli}

\begin{IEEEbiography}[{\includegraphics[width=1in,height=1.25in,clip,keepaspectratio]{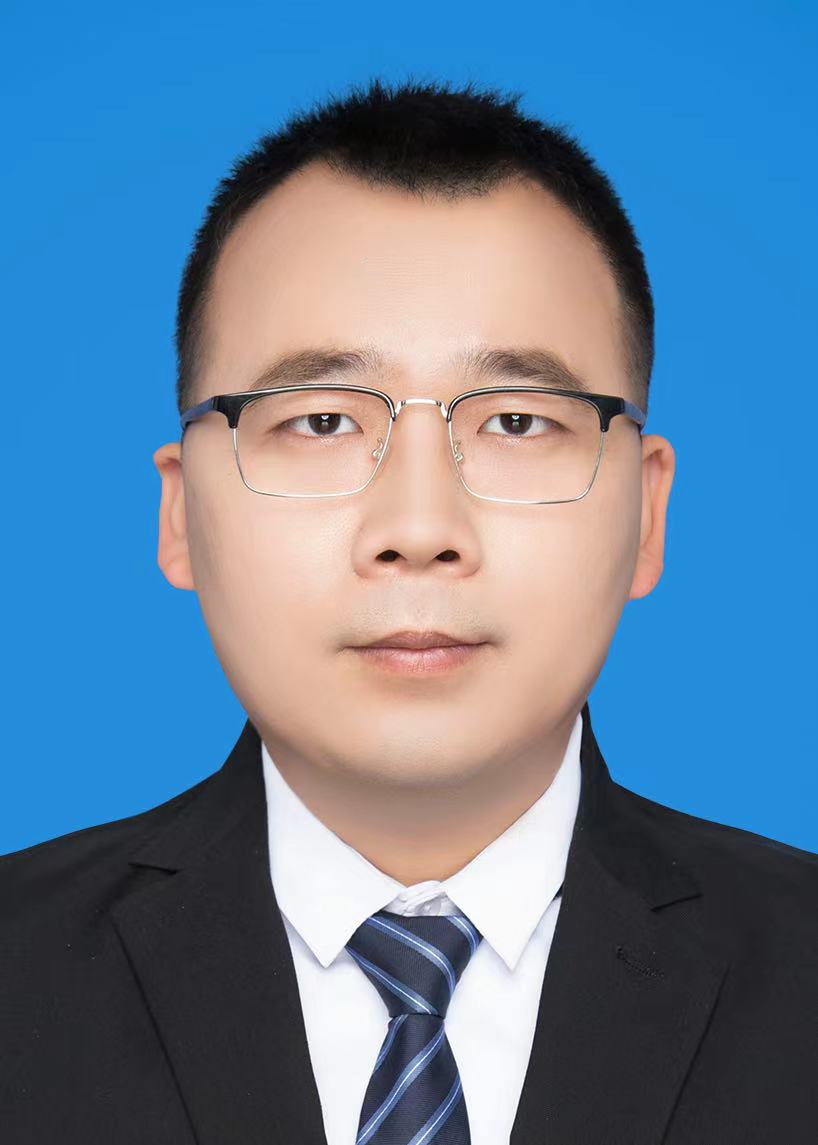}}]{Huan Gao} was born in Shandong, China. He received the B.S. degree in automation and the M.Sc. degree in control theory and control engineering from Northwestern Polytechnical University, Xi'an, Shaanxi, China, in 2011 and 2015, respectively, and the Ph.D. degree in electrical engineering from Clemson University, Clemson, SC, USA, in 2020. He is currently an Associate Professor with the School of Automation, Northwestern Polytechnical University. His research interests include decentralized optimization, cooperative control, and privacy preservation in distributed systems.
\end{IEEEbiography}

\begin{IEEEbiography}[{\includegraphics[width=1in,height=1.25in,clip,keepaspectratio]{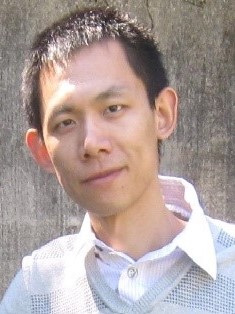}}]{Yongqiang Wang} was born in Shandong, China. He received the B.S. degree in electrical engineering and automation, the B.S. degree in computer science and technology from Xi'an Jiaotong University, Xi'an, Shaanxi, China, in 2004, and the M.Sc. and Ph.D. degrees in control science and engineering from Tsinghua University, Beijing, China, in 2009. From 2007 to 2008, he was with the University of Duisburg-Essen, Germany, as a Visiting Student. He was a Project Scientist with the University of California at Santa Barbara, Santa Barbara. He is currently an Associate Professor with the Department of Electrical and Computer Engineering, Clemson University. His current research interests include distributed control, optimization, and learning, with emphasis on privacy protection. He currently serves as an associate editor for IEEE Transactions on Control of Network systems and IEEE Transactions on Automatic Control.
\end{IEEEbiography}

\end{document}